\documentclass[reqno]{amsart}
\usepackage[printedin]{myamsart}

\usepackage{graphicx}
\input{mydef}

\providecommand{\myhbar}{h}
\providecommand{\orbit}[1]{\mathcal{O}_{#1}}
\providecommand{\uir}[1]{\rho_{#1}}
\providecommand{\oper}[1]{\mathcal{#1}}
\providecommand{\matr}[4]{{\ensuremath{ \left(\!\! \begin{array}{cc}
#1 & #2 \\ #3 & #4
\end{array}\!\!\right) }}}
\providecommand{\twovect}[2]{{\ensuremath{ \left(\!\! \begin{array}{c}
#1 \\ #2 \end{array}\!\!\right) }}}

\providecommand{\circstack}[1]{\stackrel{\scriptscriptstyle\circ}{#1}}

\newcommand{\anti}{\mathcal{A}}
\newcommand{\fock}{F_2(\uorb)}

\newcommand{\horb}{\mathcal{O}_{\hbar}}

\newcommand{\Partial}[1]{ \frac{\partial}{\partial #1} }

\newcommand{\Fracpartial}[2]{\frac{\partial #1}{\partial #2} }

\newcommand{\Diffl}[1]{\frac{d}{d #1}}

\newcommand{\Fracdiffl}[2]{\frac{d #1}{d #2}}

\newcommand{\oorb}{\mathcal{O}_0}

\newcommand{\Symp}[1]{ Sp(#1, \Space{R}{} ) }

\newcommand{\Heisn}{\Space{H}{n}}

\newcommand{\uorb}{\mathcal{O}_{h}}

\newcommand{\hilbh}{\mathcal{H}_{h}}

\newcommand{\iilbh}{\mathcal{I}_{h}}

\newcommand{\lkerh}{\mathcal{L}_{h}}

\newcommand{\lkero}{\mathcal{L}_{0}}

\newcommand{\loneh}{L_1(\Heisn)}

\newcommand{\antid}{\mathcal{A}}

\newcommand{\zerodel}{\delta(s) \delta(x) \delta(y) }

\newcommand{\zerodelsone}{\delta^{(1)}(s) \delta(x) \delta(y) }

\newcommand{\zerodelxone}{\delta(s) \delta^{(1)}(x) \delta(y) }

\newcommand{\zerodelyone}{\delta(s) \delta(x) \delta^{(1)}(y) }

\newcommand{\zerodelxtwo}{\delta(s) \delta^{(2)}(x) \delta(y) }

\newcommand{\zerodelytwo}{\delta(s) \delta(x) \delta^{(2)}(y) }

\newcommand{\fort}{\mathcal{F}}

\newcommand{\vab}{v_{(h,q,p)}}

\newcommand{\voo}{v_{(h,0,0)}}

\newcommand{\lab}{l_{(h,q,p)}}

\newcommand{\loo}{l_{(h,0,0)}}

\newcommand{\labo}{l_{(0,q,p)}}

\newcommand{\statem}{\mathcal{S}_h}

\newcommand{\statemo}{\mathcal{S}_0}

\newcommand{\ub}[3][]{\left\{\!#1\left[#2,#3\right]\!#1\right\}}

\usepackage{pstricks}
\ppnum{\href{http://arXiv.org/abs/quant-ph/0304023}
{arXiv:\texttt{quant-ph/0304023}}
}{LEEDS-PURE-MATH-2003-10}
{2003}

\providecommand{\fl}{}

\providecommand{\mailto}[1]{\href{mailto:#1}{#1}}

\begin{document}

\title[Observables and States $p$-Mechanics]{Observables and States\\ in $p$-Mechanics}
\author[A. Brodlie \and V. V. Kisil]%
{Alastair Brodlie \and \href{http://maths.leeds.ac.uk/~kisilv/}{Vladimir
    V. Kisil}}

\thanks{On leave from the Odessa University.}

\address{%
School of Mathematics\\
University of Leeds\\
Leeds LS2\,9JT\\
UK}

\email{\mailto{abrodlie@maths.leeds.ac.uk}}
\email{\mailto{kisilv@maths.leeds.ac.uk}}

\urladdr{\href{http://maths.leeds.ac.uk/~kisilv/}%
{http://maths.leeds.ac.uk/\~{}kisilv/}}

\begin{abstract}
  This is an up-to-date survey of the \(p\)-mechanical construction,
  which is a consistent physical theory suitable for a simultaneous
  description of  classical and quantum mechanics. Observables in
  \(p\)-mechanics are defined to be convolution operators on the
  Heisenberg group \(\Space{H}{n}\). Under irreducible representations
  of \(\Space{H}{n}\) the \(p\)-observables generate corresponding
  observables in classical and quantum mechanics. \(p\)-States are
  defined as positive linear functionals on \(p\)-observables. 
  It is shown that both states and observables can be realised as
  certain sets of functions/distributions on the Heisenberg group. The
  dynamical equations for both \(p\)-observables and \(p\)-states are
  provided. The construction is illustrated by the forced and unforced
  harmonic oscillators. Connections with the contextual interpretation of
  quantum mechanics are discussed.
\end{abstract}

\AMSMSC{81R05}{81R15, 22E27, 22E70, 43A65}

\maketitle

\tableofcontents
\listoffigures


\section{Introduction}
\label{sec:introduction}
The Copenhagen interpretation deals with quantum uncertainties through
the concepts of physical \emph{systems} and \emph{observers}. However
it is not correct to say that in the orthodox interpretation system
and observer are ``separated'' since they are both declared to be
meaningful only through their interaction during the
\emph{measurement} process. Effectively in most cases observer is
replaced by the \emph{state} of a physical system and a measurement
gives the probability for an observer to find the system in a
particular state.  Thus the triad of system-state-measurement forms
the starting point in any mathematical model of quantum
mechanics~\cite{vNeumann55}. Even the schemes opposing the Copenhagen
interpretation, e.g Bohmian mechanics~\cite{Bohm52a,Bohm52b} and
contextual interpretation~\cite{Khrennikov01a,Vaxjo01}, could not
escape this structure implemented in one or another way. For example,
the Bohm approach~\cite{Bohm52a,Bohm52b} may be interpreted in these
terms as follows:

\begin{center}
\noindent\begin{tabular}{||c|c||}
\hline
\hline
\textbf{Copenhagen interpretation} & \textbf{Bohmian approach} \\
\hline
\hline
a physical system governed& a hidden wave described\\
by the Sch\"odinger equation & by the master equation\\
\hline
an independent observer  & a physical particle \\
\hline
a measurement of the  & an accidental  choice of\\
system by the observer &  a trajectory by the particle \\
with probabilistic outcome & along the wave \\
\hline
\hline
\end{tabular}
\end{center}

Originally quantum mechanics was implemented through the Hamiltonian
formalism borrowed from classical mechanics. It became important
from the idealogical point of view to find a system-state-measurement
triple in the classical framework. The description of classical
mechanics in these words could be found in the introductory chapters of
almost all textbooks on quantum mechanics~\cite{Mackey63} since the
only ``king road'' into the quantum world is still going through the
classical ``pathway''. On the other hand textbooks devoted only to
classical theory~\cite{Arnold91,Goldstein80} usually ignore an
analysis of this observable-state-measurement triad as having 
little relevance in classical picture.

It is commonly accepted that observables form an
algebra both in quantum and classical mechanics: non-commutative
in the former case and commutative in the later. Then states are
defined as positive linear functionals on those algebras: in quantum
mechanics pure states are labelled by certain vectors in a Hilbert
spaces and classic pure states correspond to points of the phase
space. Measurements are represented by evaluation of observables on
particular states.

In this paper we are going to present a description of observables and
states within the framework of
\(p\)-mechanics~\cite{Kisil96a,Prezhdo-Kisil97,Kisil00a,Kisil02e,Brodlie03a}.
\(p\)-Mechanics
unifies Hamilton formulations of quantum and classical mechanics on
the grounds of the representation theory of the Heisenberg group. Therefore
our approach coincides with the traditional route in many principal
positions. On the other hand \(p\)-mechanics (being based only on the
natural properties of Heisenberg group) is in better agreement with
physical requirements and sheds an additional light on the known
results and constructions in both mechanics.
 
For example, the widespread agreement about observables being elements of an
algebra contradicts  the basic physical principle that all
measurements (and thus observables) are evaluated in certain physical
units. The multiplication of two physical quantities with different
units is natural, for example, velocity multiplied by time gives
length. However the addition of two physical quantities in different
units is unacceptable, for example, you cannot add something measured
in kilogrammes with something measured in centimetres. In
\(p\)-mechanics we don't allow addition to be freely permitted by
replacing ``algebra'' with ``symmetric space''.
For a more in depth description of this dimensional analysis see
\cite[\S\S~1.1, 1.2]{Kisil02e}.

\(p\)-Mechanical observables~\cite{Kisil96a,Kisil02e} are identified
with convolution operators on the Heisenberg group and labelled by
kernels of these convolutions. By the representations of the
Heisenberg group convolutions are transformed into well known images
of quantum (operators on a Hilbert space) and classical (functions on
the phase space) observables. Therefore it is natural to
define~\cite{Brodlie03a} \(p\)-states in line with quantum and classic
cases as positive linear functionals on the space of
\(p\)-observables. Elaboration of this approach is the main purpose of
the present paper.


\emph{The paper outline} is as follows. In the next Section we present
the representation theory of the Heisenberg group based on the
\emph{orbit method} of Kirillov~\cite{Kirillov99} and utilising
Fock--Segal--Bargmann spaces~\cite{Folland89,Howe80b}. We emphasise
the existence \emph{and} applicability of the family of
one-dimensional representations: they play for classical mechanics
exactly the same r\^ole as the infinite dimensional
representations---for quantum. In
section~\ref{sec:p-mechanics-statics} we introduce the concept of both
states and observables in \(p\)-mechanics and describe relations with
their quantum and classical counterparts. These links are provided by
the representations of the Heisenberg group and \emph{wavelet}
transforms. In subsections~\ref{sec:p-mechanical-bracket} and
\ref{sec:p-dynamic-equation} we study \(p\)-mechanical brackets and
the associated dynamic equation together with its classical and
quantum representations. While in subsections
\ref{subsect:timeevolofstates} and \ref{sect:interactpic} we describe
the time evolution of \(p\)-Mechanical states and prove that it agrees
with the time evolution of observables in doing so we exhibit the
pictures of \(p\)-dynamics. In section \ref{sect:cstates} we introduce a
system of coherent states for \(p\)-mechanics. Finally in section
\ref{sect:examplesjointp} we demonstrate the theory through the
examples of the forced and unforced harmonic oscillator.

\section{The Heisenberg Group and Its Representations}
\label{sec:preliminaries}

We start from the representation theory of the Heisenberg group
\(\Space{H}{n}\) based on the \emph{orbit method} of
Kirillov. Analysis of the unitary dual of \(\Space{H}{n}\) in
Subsection~\ref{sec:struct-topol-unit} suggests that the family of
one-dimensional representations of \(\Space{H}{n}\) forms the phase
space of a classical system. Infinite dimensional representations in a
Fock type space are described in
Subsection~\ref{sec:fock-type-spaces}.

As was mentioned in the Introduction \(p\)-mechanics does not
associate observables with an algebra since this contradicts the
physical reality. In order to emphasise that feature of
\(p\)-mechanics we supply the analysis of dimensions along our
construction.

Let \(M\) be a unit of mass, \(L\)---of length, \(T\)---of time. Then
coordinates of a point in phase space, $(q,p)$ are measured in units
\(L\) and \(ML/T\) (momentum) respectively. Derivatives
\(\frac{\partial }{\partial q}\), \(\frac{\partial }{\partial p}\),
\(\frac{\partial }{\partial t}\) of an observables with respect to
coordinates, momentum, and time are measured by \(1/L\), \(T/(ML)\),
and \(1/T\) respectively. Corresponding differentials \(dq\), \(dp\),
and \(dt\) are measured in the according units: \(L\), \(ML/T\), and
\(T\) in order to make the inner product of vectors and \(1\)-forms a
dimensionless pure number. Throughout this paper only physical
quantities of the same dimension can be added or subtracted together. 

\subsection{Representations $\Space{H}{n}$ and the Method of Orbits}
\label{sec:repr-meth-orbit}

Let \((s,x,y)\), where \(x\), \(y\in \Space{R}{n}\) and
\(s\in\Space{R}{}\), be an element of the Heisenberg group
\(\Space{H}{n}\)~\cite{Folland89,Howe80b}. We assign to \(x\) and
\(y\) components of \((s,x,y)\) physical units \(1/L\) and
  \(T/(LM)\) respectively. We chose these units so that \(qx\)
  and \(py\) are dimensionless products.

The group law on \(\Space{H}{n}\) is given as follows:
\begin{equation}
  \label{eq:H-n-group-law} (s,x,y)*(s',x',y')=(s+s'+\frac{1}{2}
  \omega(x,y;x',y'),x+x',y+y'),
\end{equation} where the non-commutativity is solely due to
\(\omega\)---the \emph{symplectic form}~\cite[\S~37]{Arnold91} on the
Euclidean space \(\Space{R}{2n}\):
\begin{equation}
  \label{eq:symplectic-form} \omega(x,y;x',y')=xy'-x'y.
\end{equation} Consequently the parameters \(s\) should be measured in
\(T/(L^2 M)\)---the product of units of \(x\) and \(y\).  The Lie
algebra \(\algebra{h}^n\) of \(\Space{H}{n}\) is spanned by a basis
\(S\), \(X_j\), \(Y_j\), \(j=1,\ldots,n\) which may be represented by
either left- or right-invariant vector fields on \(\Space{H}{n}\):
\begin{equation}
  S^{l(r)}=\pm\frac{\partial}{\partial s}, \qquad
  X_j^{l(r)}=\pm\frac{\partial}{\partial
  x_j}-\frac{y_j}{2}\frac{\partial}{\partial s}, \qquad
 Y_j^{l(r)}=\pm\frac{\partial}{\partial
 y_j}+\frac{x_j}{2}\frac{\partial}{\partial s}
  \label{eq:h-lie-algebra}
\end{equation} with the Heisenberg \emph{commutator relations}
\begin{equation}
  \label{eq:heisenberg-comm}
  [X_i^{l(r)},Y_j^{l(r)}]=\delta_{i,j}S^{l(r)}
\end{equation} all other commutators (including any between left and
right vector fields) vanish. Units to measure \(S^{l(r)}\),
\(X_j^{l(r)}\), and \(Y_j^{l(r)}\) are inverse to \(s\), \(x\),
\(y\)---i.e.  \(L^2M/T\), \(L\), and \(LM/T\) respectively---which are
obviously compatible with~\eqref{eq:heisenberg-comm}.

The exponential map \(\exp: \algebra{h}^n\rightarrow \Space{H}{n}\) is
provided by the formula:
\begin{displaymath}
  \exp: sS+\sum_{j=1}^n (x_jX_j+y_jY_j) \mapsto
  (s,x_1,\ldots,x_n,y_1,\ldots,y_n).
\end{displaymath} which respects
multiplication~\eqref{eq:H-n-group-law} and the Heisenberg
commutator relations~\eqref{eq:heisenberg-comm}. The composition of the
exponential map with representations~\eqref{eq:h-lie-algebra} of
\(\algebra{h}^n\) by the left(right)-invariant vector fields produces
the right (left) regular representation \(\lambda_{r(l)}\) of
\(\Space{H}{n}\) by right (left)
shifts. Linearised~\cite[\S~7.1]{Kirillov76} to
\(\FSpace{L}{2}(\Space{H}{n})\) they are:
\begin{equation}
  \label{eq:left-right-regular} \lambda_{r}(g): f(h) \mapsto f(hg),
  \quad \lambda_{l}(g): f(h) \mapsto f(g^{-1}h), \qquad \textrm{where
  } f(h)\in\FSpace{L}{2}(\Space{H}{n}).
\end{equation}

As any group \(\Space{H}{n}\) acts on itself by the conjugation
automorphisms \(\object[]{A}(g) h= g^{-1}hg\), which fix the unit
\(e\in \Space{H}{n}\). The differential \(\object[]{Ad}:
\algebra{h}^n\rightarrow \algebra{h}^n\) of \(\object[]{A}\) at \(e\)
is a linear map which can be differentiated again to the
representation \(\object[]{ad}\) of the Lie algebra \(\algebra{h}^n\)
by the commutator: \(\object{ad}(A): B \mapsto [B,A]\). The adjoint
space \(\algebra{h}^*_n\) of the Lie algebra \(\algebra{h}^n\) can be
realised by the left invariant first order differential forms on
\(\Space{H}{n}\). By the duality between \(\algebra{h}^n\) and
\(\algebra{h}^*_n\) the map \(\object[]{Ad}\) generates the
\emph{co-adjoint representation}~\cite[\S~15.1]{Kirillov76}
\(\object[^*]{Ad}: \algebra{h}^*_n \rightarrow \algebra{h}^*_n\):
\begin{equation}
  \label{eq:co-adjoint-rep} \object[^*]{Ad}(s,x,y): (\myhbar ,q,p)
  \mapsto (\myhbar , q+\myhbar y, p-\myhbar x), \quad \textrm{where }
  (s,x,y)\in \Space{H}{n}
\end{equation} and \((\myhbar ,q,p)\in\algebra{h}^*_n\) in
bi-orthonormal coordinates to the exponential ones on
\(\algebra{h}^n\).  These coordinates \(h\), \(q\), \(p\) should have
units of \emph{action} \(ML^2/T\), \emph{position} \(L\), and
\emph{momenta} \(LM/T\) respectively.

There are two types of orbits in~\eqref{eq:co-adjoint-rep} for
\(\object[^*]{Ad}\): the Euclidean spaces \(\Space{R}{2n}\) and single
points:
\begin{eqnarray}
  \label{eq:co-adjoint-orbits-inf} \orbit{\myhbar} & = & \{(\myhbar,
  q,p): \textrm{ for a fixed }\myhbar\neq 0 \textrm{ and all } (q,p)
  \in \Space{R}{2n}\}, \\
  \label{eq:co-adjoint-orbits-one}
  \orbit{(q,p)} & = & \{(0,q,p): \textrm{ for a fixed } (q,p)\in \Space{R}{2n}\}.
\end{eqnarray} The \emph{orbit method} of
Kirillov~\cite[\S~15]{Kirillov76}, \cite{Kirillov99} starts from the
observation that the above orbits parametrise all irreducible unitary
representations of \(\Space{H}{n}\). All representations are
\emph{induced}~\cite[\S~13]{Kirillov76} by a character
\(\chi_\myhbar(s,0,0)=e^{2\pi\rmi \myhbar s}\) of the centre of
\(\Space{H}{n}\) generated by \((\myhbar,0,0)\in\algebra{h}^*_n\) and
shifts~\eqref{eq:co-adjoint-rep} from the ``left hand side'' (i.e. by
\(g^{-1}\)) on orbits. Using~\cite[\S~13.2, Prob.~5]{Kirillov76} we
get a neat formula, which (unlike some others in the literature,
e.g. \cite[Chap.~1, (2.23)]{MTaylor86}) respects the rule that you
cannot add any two physical quantities of different units:
\begin{equation}
  \label{eq:stone-inf} \uir{\myhbar}(s,x,y): f_\myhbar (q,p) \mapsto
  e^{-2\pi\rmi(\myhbar s+qx+py)} f_\myhbar \left(q-\frac{\myhbar}{2}
  y, p+\frac{\myhbar}{2} x\right).
\end{equation} Exactly the same formula is obtained if we apply the
Fourier transform \(\hat{\ }: \FSpace{L}{2}(\Space{H}{n})\rightarrow
\FSpace{L}{2}(\algebra{h}_n^*)\) given by:
\begin{equation}
  \label{eq:fourier-transform} (\fort (\phi))
  (Y)=\hat{\phi}(Y)=\int_{\algebra{h}^n} \phi(\exp X) e^{-2\pi\rmi
  \scalar{X}{Y}}\,dX \qquad \textrm{ where } X\in\algebra{h}^n,\
  Y\in\algebra{h}_n^*
\end{equation} to the left regular
action~\eqref{eq:left-right-regular}, that is
\begin{equation} \label{eq:ftandshift}
\lambda_l (g) \fort = \fort \rho_h (g).
 \end{equation}
See~\cite[\S~2.3]{Kirillov99} for relations between the Fourier
transform~\eqref{eq:fourier-transform} and the orbit method.

The derived representation \(d\uir{\myhbar}\) of the Lie algebra
\(\algebra{h}^n\) defined on the vector
fields~\eqref{eq:h-lie-algebra} is:
\begin{equation}
  d\uir{\myhbar}(S)=-2\pi\rmi \myhbar I, \
  d\uir{\myhbar}(X_j)=\frac{\myhbar}{2} \partial_{p_j}-2\pi\rmi
    q_j I, \
  d\uir{\myhbar}(Y_j)=-\frac{\myhbar}{2} \partial_{q_j}-2\pi\rmi p_j
  I,
  \label{eq:der-repr-h-bar}
\end{equation} which clearly represent the commutation
rules~\eqref{eq:heisenberg-comm}.  The representation
\(\uir{\myhbar}\)~\eqref{eq:stone-inf} is reducible on the whole of
\(\FSpace{L}{2}(\orbit{\myhbar})\) as can be seen from the existence
of the set of ``right-invariant'', i.e. commuting
with~\eqref{eq:der-repr-h-bar}, differential operators:
\begin{equation}
  \fl  
  d\uir{\myhbar}^r(S)=2\pi\rmi \myhbar I, \quad
  d\uir{\myhbar}^r(X_j)=-\frac{\myhbar}{2} \partial_{p_j}-
    2\pi\rmi q_j I ,\quad
  d\uir{\myhbar}^r(Y_j)=\frac{\myhbar}{2} \partial_{q_j}- 2\pi\rmi p_j
  I.
  \label{eq:der-repr-h-bar-right}
\end{equation} These vectors fields represent the commutation
rules~\eqref{eq:heisenberg-comm} as well.

To obtain an irreducible representation defined
by~\eqref{eq:stone-inf} we need to restrict it to a subspace of
\(\FSpace{L}{2}(\orbit{\myhbar})\) where the operators
\eqref{eq:der-repr-h-bar-right} act as scalars, e.g. use a
\emph{polarisation} from \emph{geometric
quantisation}~\cite{Woodhouse92}.  Consider for \(\myhbar>0\) the
vector field \( -X_j+\rmi c_\rmi Y_j\) from the complexification of
\(\algebra{h}^n\), where a constant \(c_\rmi\) has the dimension
\(T/M\), the numerical value of \(c_\rmi\) in the given units can be
assumed to be \(1\). We introduce operators \(D^j_\myhbar\), \(1\leq
j\leq n\) representing vectors \(-X_j+\rmi c_\rmi Y_j\):
\begin{equation}
  \label{eq:Cauchy-Riemann}
  D^j_\myhbar =d\uir{\myhbar}^r(-X_j+\rmi c_\rmi Y_j)
  =\frac{\myhbar}{2} (\partial_{p_j}+ c_\rmi \rmi \partial_{q_j})+
  2\pi(c_\rmi p_j+ \rmi q_j) I =\myhbar \partial_{\bar{z}_j}+2\pi{z_j}
  I
\end{equation} where \( z_j = c_\rmi p_j+\rmi q_j\). For \(\myhbar<0\)
we define \(D^j_\myhbar =d\uir{\myhbar}^r(- c_\rmi Y_j+\rmi X_j)\).
Operators~\eqref{eq:Cauchy-Riemann} are used to give the following
classical result in terms of orbits:
\begin{thm}[Stone--von Neumann,
  cf. \textup{\cite[Chap.~1, \S~5]{Folland89},
  \cite[\S~18.4]{Kirillov76}}]
  \label{th:Stone-von-Neumann}
  All unitary irreducible representations of \(\Space{H}{n}\) are
  parametrised up to equivalence by two classes of
  orbits~\eqref{eq:co-adjoint-orbits-inf}
  and~\eqref{eq:co-adjoint-orbits-one} of the co-adjoint
  representation~\eqref{eq:co-adjoint-rep} in \(\algebra{h}^*_n\):
  \begin{enumerate}
  \item The infinite dimensional representations by transformation
    \(\uir{\myhbar}\)~\eqref{eq:stone-inf} for \(\myhbar \neq 0\) in
    Fock~\textup{\cite{Folland89,Howe80b}} space
    \(\FSpace{F}{2}(\orbit{\myhbar})\subset\FSpace{L}{2}(\orbit{\myhbar})\)
    of null solutions to the operators \(D^j_\myhbar\)
    \eqref{eq:Cauchy-Riemann}:
    \begin{equation}
      \label{eq:Fock-type-space}
      \FSpace{F}{2}(\orbit{\myhbar})=\{f_{\myhbar}(q,p) \in
      \FSpace{L}{2}(\orbit{\myhbar}) \such D^j_\myhbar f_{\myhbar}=0,\
       1 \leq j \leq n\}.
    \end{equation}
  \item The one-dimensional representations as multiplication by
    a constant on \(\Space{C}{}=\FSpace{L}{2}(\orbit{(q,p)})\) which
    drops out from~\eqref{eq:stone-inf} for \(\myhbar =0\):
    \begin{equation}
      \label{eq:stone-one} \uir{(q,p)}(s,x,y): c \mapsto e^{-{2\pi
      \rmi }(qx+py)}c,
    \end{equation} with the corresponding derived representation
    \begin{equation}
      \fl  d\uir{(q,p)}(S)=0, \qquad d\uir{(q,p)}(X_j)=-2\pi\rmi q_j
      ,\qquad d\uir{(q,p)}(Y_j)=-2\pi\rmi p_j.
      \label{eq:der-repr-qp}
    \end{equation}
  \end{enumerate}
\end{thm}

\subsection{Structure and Topology of the Unitary Dual of
  $\Space{H}{n}$}
\label{sec:struct-topol-unit} 
\begin{figure}[tbp]
  \begin{center}
    \includegraphics{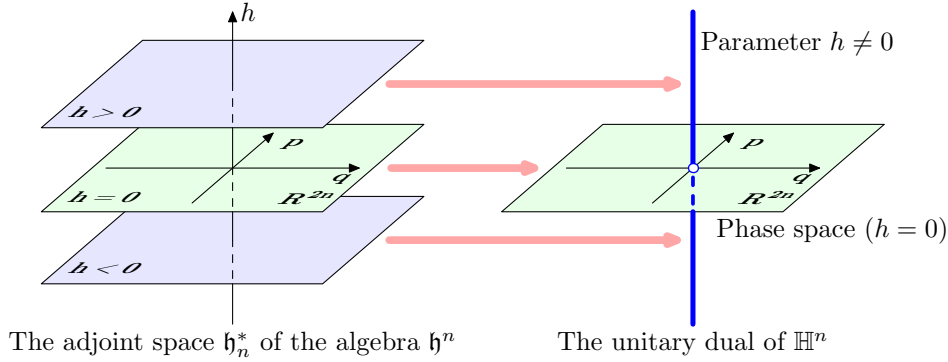}
  \end{center}
  \caption[The unitary dual object to the Heisenberg
  group]{Structure of the unitary dual object to \(\Space{H}{n}\)
    appearing from the method of orbits. The space \(\algebra{h}_n^*\)
    is sliced into ``horizontal'' hyperplanes. Planes with
    \(\myhbar\neq0\) form single orbits and correspond to different
    classes of unitary irreducible representation. The plane \(\myhbar
    =0\) is a family of one-point orbits \((0,q,p)\), which produce
    one-dimensional representations. The topology on the dual object
    is the factor topology inherited from the
    \(\algebra{h}_n^*\)~\cite[\S~2.2]{Kirillov99}.}
  \label{fi:unitary-dual}
\end{figure}

The structure of the unitary dual object
to \(\Space{H}{n}\)---the collection of all different classes of
unitary irreducible representations---as it appears from the method of
orbits is illustrated by Figure~\ref{fi:unitary-dual}, cf.
~\cite[Chap.~7, Fig.~6 and 7]{Kirillov94a}. The adjoint space
\(\algebra{h}_n^*\) is sliced into ``horizontal'' hyperplanes. A plane
with a parameter \(\myhbar\neq0\) forms a single
orbit~\eqref{eq:co-adjoint-orbits-inf} and corresponds to a particular
class of unitary irreducible representations~\eqref{eq:stone-inf}. The
plane with parameter \(\myhbar =0\) is a family of one-point orbits
\((0,q,p)\)~\eqref{eq:co-adjoint-orbits-one}, which produce
one-dimensional representations~\eqref{eq:stone-one}. The topology on
the dual object is the factor topology inherited from the adjoint
space \(\algebra{h}_n^*\) under the above identification,
see~\cite[\S~2.2]{Kirillov99}.
\begin{example}
  \label{ex:density-of-representations} A set of representations
  \(\uir{\myhbar}\)~\eqref{eq:stone-inf} with \(\myhbar\rightarrow 0\)
  is dense in the whole family of one-dimensional
  representations~\eqref{eq:stone-one}, as can be seen either from the
  Figure~\ref{fi:unitary-dual} or the analytic
  expressions~\eqref{eq:stone-inf} and~\eqref{eq:stone-one} for those
  representations.
\end{example}  Non-commutative representations \(\uir{\myhbar}\),
\(\myhbar\neq0\)~\eqref{eq:stone-inf} have been connected with quantum
mechanics from the very beginning~\cite{Folland89}, this explains, for
example, the name of the Heisenberg group. In contrast the commutative
representations~\eqref{eq:stone-one} are always neglected and only
mentioned for completeness in mathematical formulations of the
Stone--von Neumann theorem. The development of \emph{\(p\)-mechanics}
starts~\cite{Kisil96a} from the observation that the union of all
representations \(\uir{(q,p)}\), \((q,p)\in\Space{R}{2n}\) naturally
acts as the classical \emph{phase space}. The appropriateness of the
single union
\begin{equation}
  \label{eq:orbit-0} \orbit{0}=\bigcup_{(q,p)\in\Space{R}{2n}}
  \orbit{(q,p)}
\end{equation} rather than unrelated sets of disconnected orbits
manifests itself in several ways:
\begin{enumerate}
\item The topological position of \(\orbit{0}\) as the limiting case
  (cf. Example~\ref{ex:density-of-representations}) of quantum
  mechanics for \(\myhbar\rightarrow 0\) realises the
  \emph{correspondence principle} between quantum and classical
  mechanics.
\item Symplectic automorphisms of the Heisenberg group (see
  Subsection~\ref{sec:quant-from-sypl}) produce the metaplectic
  representation in quantum mechanics and \emph{transitively} acts by
  linear symplectomorphisms on the whole set \(\orbit{0}\setminus
  \{0\}\).
\item We got the Poisson brackets~\eqref{eq:Poisson} on \(\orbit{0}\)
  from the same source~\eqref{eq:u-star} which leads to the correct
  Heisenberg equation in quantum mechanics.
\end{enumerate}


Our form~\eqref{eq:stone-inf} of representations of \(\Space{H}{n}\)
given in Theorem~\ref{th:Stone-von-Neumann} has at least two following
advantages which are rarely combined together:
\begin{enumerate}
\item There is the explicit physical meaning of all entries
  in~\eqref{eq:stone-inf} as will be seen below. In contrast the
  formula (2.23) in~\cite[Chap.~1]{MTaylor86} contains terms
  \(\sqrt{\myhbar}\) (in our notations) which could be hardly
  justified from a physical point of view.
\item The one-dimensional representations~\eqref{eq:stone-one}
  explicitly correspond to the case \(\myhbar=0\)
  in~\eqref{eq:stone-inf}. The Schr\"odinger representation (the most
  used in quantum mechanics!) is handicapped in this sense: the
  transition for \(\myhbar\rightarrow 0\) from \(\uir{\myhbar}\) in
  the Sch\"odinger form to \(\uir{(q,p)}\) requires a long
  discussion~\cite[Ex.~7.11]{Kirillov94a}.
\end{enumerate}

We finish the discussion of the unitary dual of \(\Space{H}{n}\) by a
remark about negative values of \(\myhbar\). Due to its position in
the Heisenberg equation the negative value of \(\hbar\) will revert
the flow of time. Thus representations \(\uir{\myhbar}\) with
\(\myhbar<0\) seem to be suitable for a description of anti-particles
with the explicit (cf. Figure~\ref{fi:unitary-dual}) mirror symmetry
between matter and anti-matter through classical mechanics. In this
paper however we will consider only the case of \(\myhbar>0\).

\subsection{Fock Spaces $\FSpace{F}{2}(\orbit{\myhbar})$}
\label{sec:fock-type-spaces}
Our Fock type spaces~\eqref{eq:Fock-type-space} are not very
different~\cite[Ex.~4.3]{Kisil98a} from the standard Segal--Bargmann
spaces.
\begin{defn} \textup{\cite{Folland89,Howe80b}}
  The Segal--Bargmann space (with a parameter \(\myhbar>0\)) consists
  of functions on \(\Space{C}{n}\) which are holomorphic in \(z\),
  i.e. \(\partial_{\bar{z}_j} f(z)=0\), and square integrable with
  respect to the measure \(e^{-2\modulus{z}^2/\myhbar }dz\) on
  \(\Space{C}{n}\):
  \begin{displaymath}
    \int_{\Space{C}{n}} \modulus{f(z)}^2e^{-2\modulus{z}^2/\myhbar }dz
    < \infty.
  \end{displaymath}
\end{defn} Noticing the \(\partial_{\bar{z}_j}\) component in the
operator \(D^j_\myhbar\)~\eqref{eq:Cauchy-Riemann} we obviously obtain
\begin{prop}\textup{\cite{Kisil02e}}
 A function
  \(f_\myhbar(q,p)\) is in
  \(\FSpace{F}{2}(\orbit{\myhbar})\)~\eqref{eq:Fock-type-space} for
  \(\myhbar>0\) if and only if the function
  \(f_\myhbar(z)e^{\modulus{z}^2/\myhbar}\), \(z=p+\rmi q\) is in the
  classical Segal--Bargmann space.
\end{prop}

The space \(\FSpace{F}{2}(\orbit{\myhbar})\) can also be described in
the language of \emph{coherent states}, (also known as
\emph{wavelets}, \emph{matrix elements} of representation,
\emph{Berezin transform}, etc., see~\cite{AliAntGazMue}).  Since the
representation \(\uir{\myhbar}\) is irreducible any vector \(f_0\) in
\(\FSpace{F}{2}(\orbit{\myhbar})\) is \emph{cyclic},
i.e. vectors \(\uir{\myhbar}(g)f_0\) for all \(g\in G\) span the space
\(\FSpace{F}{2}(\orbit{\myhbar})\). However even if all vectors are
equally good in principle, some of them are more equal for particular
purposes. Our best option is to take
the vector in $\fock$ corresponding to the \emph{vacuum state} of the
harmonic oscillator with classical Hamiltonian $\frac{1}{2} (m
\omega^2 q^2 + \frac{1}{m}p^2)$ where $\omega$ is the constant
frequency (measured in units $\frac{1}{T}$) and $m$ is the constant
mass:
\begin{equation}
  \label{eq:vacuum} f_0(q,p)= \exp\left(-\frac{2\pi}{\myhbar}\left(
  (\omega m) q^2
    + (\omega m)^{-1} p^2\right)\right),
\end{equation} which corresponds to the minimum level of energy. Note
also that \(f_0(q,p)\) is destroyed by the \emph{annihilation
operators} (sf.~\eqref{eq:der-repr-h-bar}
and~\eqref{eq:Cauchy-Riemann}):
\begin{equation}
  \label{eq:annihilation-operator} A_\myhbar^j=d\uir{\myhbar}(
  X_j+\rmi c_\rmi Y_j)=\frac{\myhbar}{2} (\partial_{p_j}- \rmi c_\rmi
  \partial_{q_j})+2\pi(c_\rmi p_j- \rmi q_j) I.
\end{equation}

We introduce a dimensionless inner product on $\fock$ by the formula:
\begin{equation}
  \label{eq:inner-product}
  \scalar{f_1}{f_2}=\left(\frac{4}{\myhbar}\right)^{n}
  \int_{\Space{R}{2n}} f_1(q,p)\,\bar{f}_2(q,p)\,dq\,dp
\end{equation} With respect to this product the vacuum
vector~\eqref{eq:vacuum} is normalised: \(\norm{f_0}=1\).  For a
dimensionless vector $f \in \fock$ the formula defines a state
\begin{equation} \label{stateinfock}
  \scalar{A f}{f}=\left(\frac{4}{\myhbar}\right)^{n}
  \int_{\Space{R}{2n}} Af(q,p)\,\bar{f}(q,p)\,dq\,dp
\end{equation} which for any observable $A$ will give an
expectation in the units of \(A\), since the inner product is
dimensionless.The term \(\myhbar^{-n}\) in~\eqref{eq:inner-product}
not only normalises the vacuum and fixes the dimensionality of the
inner product; it is also related to the \emph{Plancherel
measure}~\cite[(1.61)]{Folland89}, \cite[Chap.~1, Th.~2.6]{MTaylor86}
on the unitary dual of \(\Space{H}{n}\).

Elements \((s,0,0)\) of the centre of \(\Space{H}{n}\) trivially act
in the representation \(\uir{\myhbar}\)~\eqref{eq:stone-inf} as
multiplication by scalars,
e.g. any function is a common eigenvector of all operators
\(\uir{\myhbar}(s,0,0)\). Thus the essential part of the operator
\(\uir{\myhbar}(s,x,y)\) is determined solely by
\((x,y)\in\Space{R}{2n}\). The \emph{coherent states} for $\fock$,
\(f_{(x,y)}(q,p)\), are ``left shifts'' of the vacuum vector
\(f_0(q,p)\) by operators~\eqref{eq:stone-inf}:
\begin{eqnarray}
  \label{eq:coherent-states} 
  \lefteqn{f_{(x,y)}(q,p) =\uir{\myhbar}(0,x,y)
  f_0(q,p)} \\ &=& \exp\left(-2\pi\rmi(qx+py)-\frac{2\pi}{\myhbar}
  \left(
      \omega m \left(q-\frac{\myhbar}{2} y\right)^ 2+(\omega
      m)^{-1}\left(p+\frac{\myhbar}{2}
      x\right)^2\right)\right). \nonumber
\end{eqnarray}

Now any function from the space \(\FSpace{F}{2}(\orbit{\myhbar})\) can
be represented~\cite[Ex.~4.3]{Kisil98a} as a linear superposition of
coherent states:
\begin{eqnarray}
  \label{eq:inv-wavelet-transform} f(q,p)=[\oper{M}_\myhbar
  \breve{f}](q,p)&=& \myhbar^{n} \int_{\Space{R}{2n}} \breve{f}(x,y)
  f_{(x,y)}(q,p)\,dx\,dy
\end{eqnarray} where \(\breve{f}(x,y)\) is the \emph{wavelet
  transform}~\cite{AliAntGazMue,Kisil98a} of \(f(q,p)\):
\begin{eqnarray}
  \label{eq:wavelet-transform} \breve{f}(x,y) =[\oper{W}_\myhbar
  f](x,y) &=&\scalar{f}{f_{(x,y)}}_{\FSpace{F}{2}(\orbit{\myhbar})}\\
  &=& \left(\frac{4}{\myhbar}\right)^{n} \int_{\Space{R}{2n}}
  f(q,p)\bar{f}_{(x,y)}(q,p)\,dq\,dp. \nonumber
\end{eqnarray} The formula~\eqref{eq:inv-wavelet-transform} can be
regarded~\cite{Kisil98a} as the \emph{inverse wavelet transform}
\(\oper{M}\) of \(\breve{f}(x,y)\).

This set of coherent states, $f_{(x,y)}$, are useful as an
overcomplete system of vectors in $\fock$ and in exhibiting relations
between \(p\)-mechanics and Berezin quantisation (subsection
\ref{sec:berezin-quantisation}). Unfortunately the "classical limits"
for $h \rightarrow 0$ of all these coherent states are functions
supported in the neighbourhood of $(0,0)$. Instead we want them to be
supported around different classical states $(q,p)$. This defect is
resolved in section \ref{sect:cstates} when we have a clearer
definition of what \(p\)-mechanical states are.

\section{$p$-Mechanics: Statics}
\label{sec:p-mechanics-statics}
We define \(p\)-mechanical observables to be convolutions on the
Heisenberg group. The next subsection describes their multiplication
and commutator as well as their quantum and classical
representations. The Berezin quantisation in the form of a wavelet
transform is considered in
subsection~\ref{sec:berezin-quantisation}. This is developed in
subsection~\ref{sec:weyl-quantisation} into a construction of
\(p\)-observables out of either quantum or classical
ones. \(p\)-Mechanical states are introduced in subsection
\ref{subsect:whatarestates}, as functionals on the set of observables,
which come in two forms: kernels and elements of a Hilbert space.

\subsection{Observables in $p$-Mechanics, Convolutions and
Commutators} \label{sec:conv-algebra-hg}

In line with the standard quantum theory we give the following
definition:
\begin{defn} \textup{\cite{Kisil96a,Kisil02e}}
  \label{de:p-observables} \emph{Observables} in \(p\)-mechanics
  (\(p\)-observables) are presented by operators on
  \(\FSpace{L}{2}(\Space{H}{n})\).
\end{defn} It is important for subsection \ref{subsect:whatarestates}
to note that as the observables are operators on a Hilbert space they
form a $C^*$-algebra \cite{Arveson76,Dixmier77}.  Actually we will
need here\footnote{More
  general operators are in use for a string-like version of
  \(p\)-mechanics, see \cite[Sect 5.2.3]{Kisil02e}.} only operators
generated by convolutions on \(\FSpace{L}{2}(\Space{H}{n})\).  Let
\(dg\) be a left invariant measure~\cite[\S~7.1]{Kirillov76} on
\(\Space{H}{n}\), which coincides with the standard Lebesgue measure
on \(\Space{R}{2n+1}\) in the exponential coordinates
\((s,x,y)\). Then a function \(B_1\) from the linear space
\(\FSpace{L}{1}(\Space{H}{n},dg)\) acts on
\(B_2\in\FSpace{L}{2}(\Space{H}{n},dg)\) by a convolution as follows:
\begin{eqnarray}
  (B_1 * B_2) (g) &=& c_\myhbar^{n+1} \int_{\Space{H}{n}} B_1(g_1)\,
  B_2(g_1^{-1}g)\,dg_1 \label{eq:de-convolution}\\ &=& c_\myhbar^{n+1}
  \int_{\Space{H}{n}} B_1(gg_1^{-1})\, B_2(g_1)\,dg_1.  \nonumber
\end{eqnarray} where the constant \(c_\myhbar\) has the value \(1\) in
the units of action. Then \(c_\myhbar^{n+1}\) has units inverse to
\(dg\). Thus the convolution \(B_1 * B_2\) is measured in units which
are the product of the units for \(B_1\) and \(B_2\). We can
alternatively write the convolution of two functions on the Heisenberg
group as
\begin{equation} \label{eq:altconvol}
(B_1*B_2) (g) = c_h^{n+1} \int_{\Heisn} B_1 (h) \lambda_l (h) dh B_2 (g)
\end{equation}
where $\lambda_l$ is as defined in equation (\ref{eq:left-right-regular}). This form of convolution is shown to be useful in subsection \ref{subsect:whatarestates}.

The composition of two convolution operators \(K_1\) and \(K_2\) with
kernels \(B_1\) and \(B_2\) has the kernel defined by the same
formula~\eqref{eq:de-convolution}. This produces \emph{inner
derivations} \(D_B\) of \(\FSpace{L}{1}(\Space{H}{n})\) by the
\emph{commutator}:
\begin{eqnarray}
  D_B: f \mapsto [B,f]&=&B*f-f*B \nonumber \\ &=& c_\myhbar^{n+1}
  \int_{\Space{H}{n}} B(g_1)\left(
  f(g_1^{-1}g)-f(gg_1^{-1})\right)\,dg_1.
  \label{eq:commutator}
\end{eqnarray} Since we only consider observables which are
convolutions on \(\Space{H}{n}\) we can extend a unitary
representation \(\uir{\myhbar} \) of \(\Space{H}{n}\) to a
\(*\)-representation of \(\FSpace{L}{1}(\Space{H}{n} ,dg)\) by the
formula:
\begin{eqnarray}
  \fl
\lefteqn{  [\uir{\myhbar} (B)f](q,p)
  = c_\myhbar^{n+1} \int_{\Space{H}{n}} B(g)\uir{\myhbar}
  (g)f(q,p)\,dg}  \label{eq:rho-extended-to-L1} \\
  &=& c_\myhbar^{n} \int_{\Space{R}{2n}}
  \left(c_\myhbar\int_{\Space{R}{}} B(s,x,y)
    e^{-2\pi\rmi \myhbar s}\,ds \right)
  e^{- 2\pi\rmi (qx+py)} f
  \left(q-\frac{\myhbar}{2} y, p+\frac{\myhbar}{2} x\right) \,dx\,dy.
  \nonumber
\end{eqnarray} The last formula in the Schr\"odinger representation
defines for \(\myhbar \neq 0\) a \emph{pseudodifferential
  operator}~\cite{Folland89,Howe80b,Shubin01} on
\(\FSpace{L}{2}(\Space{R}{n})\)~\eqref{eq:Fock-type-space}, which are
known to be \emph{quantum observables} in the Weyl quantisation. For
representations \(\uir{(q,p)}\)~\eqref{eq:stone-one} the expression
analogous to~\eqref{eq:rho-extended-to-L1} defines an operator of
multiplication on \(\orbit{0}\)~\eqref{eq:orbit-0} by the Fourier
transform of \(B(s,x,y)\):
\begin{equation}
  \label{eq:classical-observables}
  \uir{(q,p)}(B)=\hat{B}\left(0,{q},{p}\right) = c_\myhbar^{n+1}
  \int_{\Space{H}{n}} B(s,x,y)e^{-{2\pi\rmi}(qx+py)} \,ds\,dx\,dy,
\end{equation} where the direct \(\hat{\ }\) and inverse \(\check{\
}\) Fourier transforms are defined by the formulae:
\begin{displaymath}
  \hat{f}(v)=\int_{\Space{R}{m}} f(u) e^{-2\pi\rmi uv}\,du \qquad
  \textrm{ and } \qquad f(u)=(\hat{f})\check{\
  }(u)=\int_{\Space{R}{m}} \hat{f}(v) e^{2\pi\rmi vu}\,dv.
\end{displaymath} For reasons discussed in
subsections~\ref{sec:struct-topol-unit}
and~\ref{sec:p-mechanical-bracket} we regard the
functions~\eqref{eq:classical-observables} on \(\orbit{0}\) as
\emph{classical observables}.  Again both the representations
\(\uir{\myhbar} (B)\) and \(\uir{(q,p)}(B)\) are measured in the same
units as the function \(B\).

From~\eqref{eq:rho-extended-to-L1} it follows that \(\uir{\myhbar}
(B)\) for a fixed \(\myhbar \neq 0\) depends only from
\(\hat{B}_s(\myhbar,x,y)\)---the partial Fourier transform
\(s\rightarrow \myhbar\) of \(B(s,x,y)\). Then the representation of
the composition of two convolutions depends only from
\begin{eqnarray}
  \fl
  (B'*B)\hat{_s}
  &=& 
  c_\myhbar \int_{\Space{R}{}}e^{ -2\pi\rmi \myhbar s}\,
  c_\myhbar^{n+1} \int_{\Space{H}{n}}B'(s',x',y')
 \label{eq:almost-star-product}\\
 &&\quad \qquad \times B(s-s'+\frac{1}{2}
 (xy'-yx'),x-x',y-y')\,ds'dx'dy'ds \nonumber \\
  &=& c_\myhbar^{n} \int_{\Space{R}{2n}} e^{ {\pi\rmi \myhbar}{}
   (xy'-yx')} \hat{B}'_s(\myhbar ,x',y')
 \hat{B}_s(\myhbar ,x-x',y-y')\,dx'dy'. \nonumber
\end{eqnarray} Note that if we apply the Fourier transform
\((x,y)\rightarrow (q,p)\) to the last expression
in~\eqref{eq:almost-star-product} then we get the \emph{star product}
of \(\hat{B}'\) and \(\hat{B}\) known in \emph{deformation}
quantisation, cf. \cite[(9)--(13)]{Zachos02a}.  Consequently the
representation \(\uir{\myhbar}([B',B])\) of
commutator~\eqref{eq:commutator} depends only from:
\begin{eqnarray}
  \fl
\label{eq:repres-commutator}
  \lefteqn{[B',B]\hat{_s}
    =
    c_\myhbar^{n}\int_{\Space{R}{2n}}
    \left(e^{ {\rmi\pi}{} \myhbar (xy'-yx')}-e^{- {\rmi\pi}{} \myhbar
        (xy'-yx')}\right)}
  \qquad \\
  && \qquad \qquad \qquad \times 
  \hat{B}'_s(-\myhbar ,x',y')
  \hat{B}_s(-\myhbar ,x-x',y-y') \,dx'dy'\nonumber \\
 \qquad&=& 2 \rmi c_\myhbar^n \int_{\Space{R}{2n}}\! 
 \sin\left({\pi\myhbar}
   (xy'-yx')\right)
 \hat{B}'_s(\myhbar ,x',y') \hat{B}_s(\myhbar
 ,x-x',y-y')\,dx'dy'.\nonumber
\end{eqnarray} The integral~\eqref{eq:repres-commutator} turns out to
be equivalent to the \emph{Moyal brackets}~\cite{Zachos02a} for the
(full) Fourier transforms of \(B'\) and \(B\). It is commonly accepted
that the method of orbits is the mathematical side of \emph{geometric}
quantisation~\cite{Woodhouse92}. Our derivation of the Moyal brackets
in terms of orbits shows that deformation and geometric quantisations
are closely connected and both are not very far from the original
quantisation of Heisenberg and Schr\"odinger. Yet one more close
relative can be identified as the Berezin
quantisation~\cite{Berezin75}, see the next subsection.

\begin{rem}
  \label{re:triv-commutator} The
  expression~\eqref{eq:repres-commutator} vanishes for \(\myhbar=0\)
  as can be expected from the commutativity of
  representations~\eqref{eq:stone-one}. Thus it does not produce
  anything interesting on \(\orbit{0}\), that supports the common
  negligence to this set.
\end{rem}

Summing up, \(p\)-mechanical observables, i.e. convolutions on
\(\FSpace{L}{2}(\Space{H}{n})\) are transformed
\begin{enumerate}
\item by representations \(\uir{\myhbar}\)~\eqref{eq:stone-inf}
  into quantum observables~\eqref{eq:rho-extended-to-L1} with the
  Moyal bracket~\eqref{eq:repres-commutator} between them;
\item by representations \(\uir{(q,p)}\)~\eqref{eq:stone-one} into
    classical observables~\eqref{eq:classical-observables}.
\end{enumerate} We haven't got a meaningful bracket on the set of
classical observables yet, this will be done in
Section~\ref{sec:p-mechanical-bracket}.

\subsection{Berezin Quantisation and Wavelet Transform}
\label{sec:berezin-quantisation}

There is the following construction, known as the \emph{Berezin
quantisation}, which allows us to assign a function to an operator
(observable) and an operator to a function. The scheme is based on the
construction of \emph{coherent states}, which can be derived from
different sources~\cite{Klauder94b,Perelomov86}. We prefer the
group-theoretic origin of Perelomov coherent states~\cite{Perelomov86}
in this section we use the coherent states in $\fock$ defined in
equation~\eqref{eq:coherent-states}. Later in this paper we construct
a more general system of coherent states independent of a Hilbert
space. Following~\cite{Berezin75} we introduce a \emph{covariant}
symbol \({a}(g)\) of an operator \(A\) on
\(\FSpace{F}{2}(\orbit{\myhbar})\) by the simple expression:
\begin{equation}
  \label{eq:covariant} {a}(g)=\scalar{A f_g}{f_g},
\end{equation}
i.e. we get a map from the linear space of operators on
\(\FSpace{F}{2}(\orbit{\myhbar})\) to a linear space of functions on
\(\Space{H}{n}\).  A map in the opposite direction assigns to a
function \(\breve{a}(g)\) on \(\Space{H}{n}\) a linear operator \(A\)
on \(\FSpace{F}{2}(\orbit{\myhbar})\) by the formula
\begin{equation}
  \label{eq:contravariant} A=c_\myhbar^{n+1}\int_{\Space{H}{n}}
  \circstack{a}(g)P_g\,dg, \qquad \textrm{ where \(P_g\) is
    the projection } P_g
  f=\scalar{f}{f_g}f_g.
\end{equation} The function \(\circstack{a}(g)\) is called the
\emph{contravariant} symbol of the operator
\(A\)~\eqref{eq:contravariant}.

The co- and contravariant symbols of operators are defined through the
coherent states, in fact both types of symbols are
realisations~\cite[\S~3.1]{Kisil98a} of the
direct~\eqref{eq:wavelet-transform} and
inverse~\eqref{eq:inv-wavelet-transform} wavelet transforms.  Let us
define a representation \(\uir{b\myhbar}\) of the group \(\Space{H}{n}
\times \Space{H}{n} \) in the space
\(\oper{B}(\FSpace{F}{2}(\orbit{\myhbar}))\) of operators on
\(\FSpace{F}{2}(\orbit{\myhbar})\) by the formula:
\begin{equation}
  \label{eq:bi-representation} \uir{b\myhbar}(g_1,g_2): A \mapsto
  \uir{\myhbar}(g_1^{-1})A\uir{\myhbar}(g_2), \qquad \textrm{ where }
  g_1,g_2\in\Space{H}{n}.
\end{equation} According to the scheme from~\cite{Kisil98a} for any
state \(l_0\) on \(\oper{B}(\FSpace{F}{2}(\orbit{\myhbar}))\) we get a
wavelet transform \(\oper{W}_{l_0}:
\oper{B}(\FSpace{F}{2}(\orbit{\myhbar}))\rightarrow
\FSpace{C}{}(\Space{H}{n}\times\Space{H}{n}) \):
\begin{equation}
  \label{eq:wavelet-transform-operators} \oper{W}_{l_0}: A \mapsto
  \breve{a}(g_1,g_2)= \scalar{\uir{b\myhbar}(g_1,g_2)A}{l_0}.
\end{equation}

The important particular case is given by \(l_0\) defined through the
vacuum vector \(f_0\)~\eqref{eq:vacuum} by the formula
\(\scalar{A}{l_0}=\scalar{Af_0}{f_0}\). Then the wavelet
transform~\eqref{eq:wavelet-transform-operators} produces the
\emph{covariant presymbol} \(\breve{a}(g_1,g_2)\) of operator
\(A\). Its restriction \(a(g)=\breve{a}(g,g)\) to the diagonal \(D\)
of \(\Space{H}{n}\times\Space{H}{n}\) is exactly~\cite{Kisil98a} the
Berezin covariant symbol~\eqref{eq:covariant} of~\(A\). Such a
restriction to the diagonal is done without a loss of information due
to holomorphic properties of \(\breve{a}(g_1,g_2)\)~\cite{Berezin72}.

Another important example of the state \(l_0\) is given by the trace:
\begin{equation}
  \label{eq:trace} \scalar{A}{l_0}=\object{tr}A =\myhbar^n
  \int_{\Space{R}{2n}} \scalar[\FSpace{F}{2}(\orbit{\myhbar})]{A
    f_{(x,y)}}{f_{(x,y)}}\,dx\,dy,
\end{equation} where coherent states \(f_{(x,y)}\) are again defined
in~\eqref{eq:coherent-states}. The operators \(\uir{b\myhbar}(g,g)\)
from the diagonal \(D\) of \(\Space{H}{n}\times\Space{H}{n}\)
trivially act on the wavelet
transform~\eqref{eq:wavelet-transform-operators} generated by the
trace~\eqref{eq:trace} since the trace is invariant under
\(\uir{b\myhbar}(g,g)\). According to the general scheme we can
consider the \emph{reduced wavelet transform}~\cite{Kisil98a} on the
homogeneous space \(\Space{H}{n}\times\Space{H}{n}/D\) instead of the
entire group \(\Space{H}{n}\times\Space{H}{n}\). The space
\(\Space{H}{n}\times\Space{H}{n}/D\) is isomorphic to \(\Space{H}{n}\)
with an embedding \(\Space{H}{n} \rightarrow \Space{H}{n} \times
\Space{H}{n}\) given by \(g \mapsto (g;0)\). Furthermore the centre
\(Z\) of \(\Space{H}{n}\) acts trivially in the representation
\(\uir{b\myhbar}\) as usual. Thus the only essential part of
\(\Space{H}{n}\times\Space{H}{n}/D\) in the wavelet transform is the
homogeneous space \(\Omega=\Space{H}{n}/Z\). A Borel section
\(\mathbf{s}: \Omega \rightarrow \Space{H}{n}\times\Space{H}{n} \) in
the principal bundle \(G \rightarrow \Omega\) can be defined as
\(\mathbf{s}(x,y)\mapsto ((0,x,y);(0,0,0))\). We got the reduced
realisation \(\oper{W}_r\) of the wavelet
transform~\eqref{eq:wavelet-transform-operators} in the form:
\begin{eqnarray}
  \oper{W}_r: A \mapsto
  \breve{a}_r(x,y)&=&\scalar{\uir{b\myhbar}(\mathbf{s}(x,y)) A}{l_0}
  \nonumber\\ &=& \object{tr}(\uir{\myhbar}((0,x,y)^{-1}) A)
   \label{eq:inversion-trace} \\
  &=& \myhbar^{n} \int_{\Space{R}{2n}}
  \scalar[\FSpace{F}{2}(\orbit{\myhbar})]{
    \uir{\myhbar}((0,x,y)^{-1})A f_{(x',y')}}{f_{(x',y')}}\,dx'dy'
  \nonumber \\ &=& \myhbar^{n} \int_{\Space{R}{2n}}
  \scalar[\FSpace{F}{2}(\orbit{\myhbar})]{A
    f_{(x',y')}}{f_{(x,y)*(x',y')}}\,dx'dy'.
  \label{eq:inversion-formula}
\end{eqnarray} The formula~\eqref{eq:inversion-trace} is the principal
ingredient of the \emph{inversion formula} for the Heisenberg
group~\cite[Chap.~1, (1.60)]{Folland89}, \cite[Chap.~1,
Th.~2.7]{MTaylor86}, which reconstructs the kernel of convolution
\(B(g)\) out of operators \(\uir{\myhbar}(B)\). Therefore if we define
the mother wavelet to be the identity operator \(I\) the inverse
wavelet transform (cf.~\eqref{eq:inv-wavelet-transform}) will be
\begin{eqnarray}
  \label{eq:inverse-wavelet-convolution} \oper{M}_r \breve{a}_r &=&
  \myhbar^n\int_{\Space{R}{2n}}
  a(x,y)\uir{b\myhbar}(\mathbf{s}((0,x,y)^{-1})) I\,dx\,dy\\ &=&
  \myhbar^n \int_{\Space{R}{2n}} a(x,y)\uir{\myhbar}(0,x,y)
  \,dx\,dy\nonumber.
\end{eqnarray} The inversion formula for \(\Space{H}{n}\) insures that
\begin{prop}\textup{\cite{Kisil02e}}
  The composition \(\oper{M}_r\circ\oper{W}_r\) is the identity map on
  the representations \(\uir{\myhbar}(B)\) of convolution operators on
  \(\orbit{\myhbar}\).
\end{prop} 
\begin{example}
  \label{ex:wavelet-transform-operator} The wavelet transform
  \(\oper{W}_r\)~\eqref{eq:inversion-formula} applied to the quantum
  coordinate \(Q=d\uir{\myhbar}(X)\), momentum \(P=d\uir{\myhbar}(Y)\)
  (see~\eqref{eq:der-repr-h-bar}), and the energy function of the
  harmonic oscillator \((m\omega^2 Q^2+\frac{1}{m} P^2)/2\) produces
  the distributions on \(\Space{R}{2n}\):
  \begin{eqnarray*}
    Q & \mapsto & \frac{1}{2\pi \rmi}\delta^{(1)}(x)\delta(y),\\%
    \label{eq:wavelet-Q}\\ P & \mapsto & \frac{1}{2\pi
    \rmi}\delta(x)\delta^{(1)}(y),\\
    \frac{1}{2}\left(m\omega^2 Q^2+\frac{1}{m} P^2\right) & \mapsto &
    -\frac{1}{8\pi^2} \left(m\omega^2 \delta^{(2)}(x)\delta(y)
      +\frac{1}{m} \delta(x)\delta^{(2)}(y)\right),
      \\
  \end{eqnarray*} where \(\delta^{(1)}\) and \(\delta^{(2)}\) are the
  first and second derivatives of the Dirac \emph{delta function}
  \(\delta\) respectively. We will use them later in
  Example~\ref{ex:quantum-to-p-mechanics}.
\end{example}

\subsection{From Classical and Quantum Observables to $p$-Mechanics}
\label{sec:weyl-quantisation}

It is commonly accepted that we can not deal with quantum mechanics
directly and thus classical dynamics serve as an unavoidable
intermediate step. The passage from classical observables to quantum
ones---known as a \emph{quantisation}---is a huge field with many
concurring approaches (geometric, deformation, Weyl, Berezin,
etc. quantisations) each having its own merits and demerits. Similarly
one has to construct \(p\)-mechanical observables starting from
classical or quantum ones by some procedure (should it be named
``\(p\)-mechanisation''?), which we are about to describe now.

The transition from a \(p\)-mechanical observable to a classical one
is given by the formula~\eqref{eq:classical-observables}, which in
turn is a realisation of the inverse wavelet
transform~\eqref{eq:inv-wavelet-transform}:
\begin{equation}
  \label{eq:classical-observables-1}
  \uir{(q,p)}B=\hat{B}\left(0,{q},{p}\right)
  =c_\myhbar^{n+1}\int_{\Space{H}{n}} B(s,x,y)e^{-{2\pi\rmi}(qx+py)}
  \,ds\,dx\,dy.
\end{equation}

Just like in the case of quantisation the classical image
\(\uir{(q,p)}(B)\)~\eqref{eq:classical-observables-1} contains only
partial information about a \(p\)-observable \(B\) unless we make some
additional assumptions.  Let us start from a classical observable
\(c(q,p)\) and try to construct the corresponding \(p\)-observable.
As follows from general considerations (see~\cite{Kisil98a} and
Section~\ref{sec:fock-type-spaces}) we can partially invert
formula~\eqref{eq:classical-observables-1} by the wavelet
transform~\eqref{eq:wavelet-transform}:
\begin{equation}
  \label{eq:inv-classical-observables} \check{c}(x,y) = [\oper{W}_{0}
  c] (x,y)= \scalar{c v_{(0,0)}}{v_{(x,y)}} =c_\myhbar^n
  \int_{\Space{R}{2n}} c(q,p)e^{{2\pi\rmi}(qx+py)} \,dq\,dp,
\end{equation} where
\(v_{(x,y)}=\uir{(q,p)}v_{(0,0)}=e^{-2\pi\rmi(qx+py)}\).

However the function
\(\check{c}(x,y)\)~\eqref{eq:inv-classical-observables} is not defined
on the entire of \(\Space{H}{n}\). The natural domain of
\(\check{c}(x,y)\) according to the construction of the reduced
wavelet transform~\cite{Kisil98a} is the homogeneous space
\(\Omega=G/Z\), where \(G=\Space{H}{n}\) and \(Z\) is its normal
subgroup of central elements \((s,0,0)\).  Let \(\mathbf{s}: \Omega
\rightarrow G\) be a Borel section in the principal bundle \(G
\rightarrow \Omega\), which is used in the construction of induced
representations, see~\cite[\S~13.1]{Kirillov76}. For the Heisenberg
group~\cite[Ex.~4.3]{Kisil98a} it can be simply defined as
\(\mathbf{s}: (x,y)\in\Omega\mapsto (0,x,y)\in\Space{H}{n} \). One can
naturally transfer functions from \(\Omega\) to the image
\(\mathbf{s}(\Omega)\) of the map \(\mathbf{s}\) in \(G\). However the
range \(\mathbf{s}(\Omega)\) of \(\mathbf{s}\) has often (particularly
for \(\Space{H}{n}\)) a zero Haar measure in \(G\). Probably two
simplest possible ways out are:
\begin{enumerate}
\item To increase the ``weight'' of function \(\tilde{c}(s,x,y)\)
  vanishing outside of the range \(\mathbf{s}(\Omega)\) of
  \(\mathbf{s}\) by a suitable Dirac delta function on the subgroup
  \(Z\). For the Heisenberg group this can be done, for example, by
  the map:
  \begin{equation}
    \label{eq:p-mechanisation} \oper{E}: \check{c}(x,y)\mapsto
    \tilde{c}(s,x,y)=\delta(s)\check{c}(x,y),
  \end{equation} where \(\check{c}(x,y)\) is given by the inverse
  wavelet (Fourier) transform~\eqref{eq:inv-classical-observables}. As
  we will see in Proposition~\ref{prop:Weyl-quantisation-decomposed}
  this is related to the Weyl quantisation and the Moyal brackets.
\item
  To extend the function \(\check{c}(x,y)\) to the entire group \(G\)
  by a tensor product with a suitable function on \(Z\), for example
  \(e^{-s^2}\):
  \begin{displaymath}
    \check{c}(x,y)\mapsto \tilde{c}(s,x,y)=e^{-s^2}\check{c}(x,y).
  \end{displaymath} In order to get the \emph{correspondence
  principle} between classical and quantum mechanics
  (cf. Example~\ref{ex:density-of-representations}) the function on
  \(Z\) has to satisfy some additional requirements. For
  \(\Space{H}{n}\) it should vanish for \(s\rightarrow\pm\infty\),
  which is fulfilled for both \(e^{-s^2}\) and \(\delta(s)\) from the
  previous item. In this way we get infinitely many essentially
  different quantisations with non-equivalent \emph{deformed} Moyal
  brackets between observables.
\end{enumerate} There are other more complicated possibilities not
mentioned here, which can be of some use if some additional
information or assumptions are used to extend functions from
\(\Omega\) to \(G\).  We will focus here only on the first
``minimalistic'' approach from the two listed above.

\begin{example}
  \label{ex:harmonic-oscillator-energy} The composition of the wavelet
  transform \(\oper{W}_0\)~\eqref{eq:inv-classical-observables} and
  the map \(\oper{E}\)~\eqref{eq:p-mechanisation} applied to the
  classical coordinate, momentum, and the energy function of the
  harmonic oscillator produces the distributions on \(\Space{H}{n}\):
  \begin{eqnarray}
    q & \mapsto & \frac{1}{2\pi
    \rmi}\delta(s)\delta^{(1)}(x)\delta(y), \label{eq:p-mech-q}\\ p &
    \mapsto & \frac{1}{2\pi
    \rmi}\delta(s)\delta(x)\delta^{(1)}(y),\label{eq:p-mech-p}\\
    \frac{1}{2}\left(m\omega^2 q^2+\frac{1}{m} p^2\right) & \mapsto &
    -\frac{1}{8\pi^2} \left(m\omega^2
    \delta(s)\delta^{(2)}(x)\delta(y)
      +\frac{1}{m} \delta(s)\delta(x)\delta^{(2)}(y)\right),
      \label{eq:p-mech-q2+p2}.
  \end{eqnarray}
We will use the notation $X=\frac{1}{2\pi \rmi} \zerodelxone$ and
$Y=\frac{1}{2\pi \rmi} \zerodelyone$, these distributions are used
later in Section~\ref{sect:examplesjointp}.
\end{example}

\begin{figure}[tbp]
  \begin{center}
    \includegraphics{ubracket-unit.15}
  \end{center}
    \caption[Quantisation and wavelet transforms]{The relations
      between:\\
       \(\oper{Q}_\myhbar\)---the Weyl quantisation from classical
       mechanics to quantum;\\ \(\oper{C}_{\myhbar\rightarrow
       0}\)---the classical limit \(\myhbar\rightarrow 0\) of quantum
       mechanics;\\ \(\uir{\myhbar}\) and \(\uir{(q,p)}\)---unitary
       representations of Heisenberg group \(\Space{H}{n}\); \\
       \(\oper{W}_r\) and \(\oper{W}_0\)---wavelet
      transforms defined in~\eqref{eq:inversion-trace}
      and~\eqref{eq:inv-classical-observables};\\
      \(\oper{E}\)---extension
       of functions from \(\Omega=\Space{H}{n}/Z\) to
      the whole group \(\Space{H}{n}\).\\ Note the relations
      \(\oper{Q}_\myhbar =\uir{\myhbar}\circ\oper{E}\circ\oper{W}_0\)
      and \(C_{\myhbar\rightarrow 0} =
      \uir{(q,p)}\circ\oper{E}\circ\oper{W}_r\).}
    \label{fi:quant-class-p-observ}
\end{figure}

If we apply the representation
\(\uir{\myhbar}\)~\eqref{eq:rho-extended-to-L1} to the function
\(\tilde{c}(s,x,y)\)~\eqref{eq:p-mechanisation} we will get the
operator on \(\FSpace{F}{2}(\orbit{\myhbar})\):
\begin{eqnarray}
  \oper{Q}_\myhbar (c)&=&c_\myhbar^{n+1} \int_{\Space{H}{n}}
  \tilde{c}(s,x,y)\uir{\myhbar}(s,x,y)\,ds\,dx\,dy \nonumber \\
  &=&c_\myhbar^{n} \int_{\Space{R}{2n}} \check{c}(x,y)e^{x \cdot
  d\uir{\myhbar}(X)+
    y \cdot d\uir{\myhbar}(Y) }\,dx\,dy, \label{eq:weyl-calculus}
\end{eqnarray} where the last expression is exactly the \emph{Weyl
quantisation} (the \emph{Weyl
correspondence}~\cite[\S~2.1]{Folland89}) if the Schr\"odinger
realisation with \(d\uir{\myhbar}(X)=q\) and
\(d\uir{\myhbar}(Y)=\rmi\myhbar \partial_q\) on
\(\FSpace{L}{2}(\Space{R}{n})\) is chosen for \(\uir{\myhbar}\). Thus
we demonstrate that

\begin{prop} \textup{\cite{Kisil02e}}
  \label{prop:Weyl-quantisation-decomposed} The \emph{Weyl
  quantisation} \(\oper{Q}_\myhbar\)~\eqref{eq:weyl-calculus} is the
  composition of the wavelet
  transform~\eqref{eq:inv-classical-observables}, the extension
  \(\oper{E}\)~\eqref{eq:p-mechanisation}, and the representation
  \(\uir{\myhbar}\)~\eqref{eq:stone-inf}:
  \begin{equation}
    \label{eq:Weyl-quantisation-decomposition} \oper{Q}_\myhbar
    =\uir{\myhbar}\circ\oper{E}\circ\oper{W}_0.
  \end{equation}
\end{prop}

A similar construction can be carried out if we have a quantum
observable \(A\) and wish to recover the related \(p\)-mechanical
object. The wavelet transform
\(\oper{W}_r\)~\eqref{eq:inversion-trace} maps \(A\) into the function
\(a(x,y)\) defined on \(\Omega\) and we again face the problem of
extending \(a(x,y)\) to the entire group \(\Space{H}{n}\). It will be
solved as in the classical case by a tensor product with the delta
function \(\delta(s)\). We get the following formula:
\begin{displaymath}
  A \mapsto a(s,x,y)=\oper{E}\circ\oper{W}_r (A)=\myhbar^{n}\delta(s)
  \int_{\Space{R}{2n}} \scalar[\FSpace{F}{2}(\orbit{\myhbar})]{A
    v_{(x',y')}}{v_{(x,y)\cdot(x',y')}}\,dx'dy'.
\end{displaymath} We can apply to this function \(a(s,x,y)\) the
representation \(\uir{(q,p)}\) and obtain a classical observable
\(\uir{(q,p)}(a)\). For a reasonable quantum observable \(A\) its
classical image \(\uir{(q,p)}\circ\oper{E}\circ\oper{W}_r (A)\) will
coincide with its classical limit \(C_{\myhbar\rightarrow
  0}A\):
\begin{equation}
  \label{eq:classical-limit-decomposition} C_{\myhbar\rightarrow 0} =
  \uir{(q,p)}\circ\oper{E}\circ\oper{W}_r,
\end{equation} which is expressed here through integral
transformations and does not explicitly use the limit
\(\myhbar\rightarrow 0\).  Figure~\ref{fi:quant-class-p-observ}
illustrates various transformations between quantum, classical, and
\(p\)-observables. Besides the mentioned
decompositions~\eqref{eq:Weyl-quantisation-decomposition}
and~\eqref{eq:classical-limit-decomposition} there are presentations
of identity maps on classical and quantum spaces correspondingly:
\begin{displaymath}
\oper{I}_c=\uir{(q,p)}\circ\oper{E}\circ\oper{W}_0, \qquad
\oper{I}_\myhbar=\uir{\myhbar}\circ\oper{E}\circ\oper{W}_\myhbar.
\end{displaymath}
\begin{example}
  \label{ex:quantum-to-p-mechanics} The wavelet transform
  \(\oper{W}_r\) applied to the quantum coordinate \(Q\), momentum
  \(P\), and the energy function of the harmonic oscillator \((m
  \omega^2Q^2+\frac{1}{m}P^2)/2\) was calculated in
  Example~\ref{ex:wavelet-transform-operator}. A composition with the
  above map \(\oper{E}\) yields the distributions:
  \begin{eqnarray*}
    Q & \mapsto & \frac{1}{2\pi\rmi}\delta(s)\delta^{(1)}(x)\delta(y),
    \label{eq:p-mech-Q}\\ P & \mapsto &
    \frac{1}{2\pi\rmi}\delta(s)\delta^{(1)}(x)\delta(y)
    ,\label{eq:p-mech-P}\\ \frac{1}{2}\left(m\omega^2 Q^2+\frac{1}{m}
    P^2\right) & \mapsto & -\frac{1}{2\pi^2} \left(m\omega^2
    \delta(s)\delta^{(2)}(x)\delta(y)
      +\frac{1}{m} \delta(s)\delta(x)\delta^{(2)}(y)\right),
      \label{eq:p-mech-Q2+P2}
  \end{eqnarray*} which are exactly the same as in the
  Example~\ref{ex:harmonic-oscillator-energy}.
\end{example}

\subsection{$p$-Mechanical States} \label{subsect:whatarestates}

In this subsection we introduce states to \(p\)-mechanics --- these
are positive linear functionals on the \(C^*\)-algebra
\cite{Arveson76,Dixmier77} of \(p\)-mechanical observables (cf.
subsection \ref{sec:conv-algebra-hg}). According to the GNS construction
for a general
\(C^*\)-algebra~\cite[\S~1.6]{Arveson76}
\begin{itemize}
\item an arbitrary state could be decomposed as a linear combination of
  the pure states; and 
\item the \emph{pure} states correspond to
  irreducible representations.
\end{itemize}
Since irreducible representations of  \(\FSpace{L}{1}(\Space{H}{n})\) are
given by Theorem~\ref{th:Stone-von-Neumann} and are associated in
\(p\)-mechanics with quantum and classical pictures then the pure
states in \(p\)-mechanics also corresponds to quantum and classical
states.  We give here several equivalent descriptions of these states.

For each $h \neq 0$ (the
quantum case) we give two equivalent forms of states: the first form
we give is as elements of a Hilbert space, the second is as
integration with an appropriate kernel. For $h=0$ (the classical case)
we provide only one form of states, that is as integration with an
appropriate kernel since the second one is not essentially different
from the former. 

\begin{defn}
\textup{\cite{Brodlie03a}}
The Hilbert space $\hilbh$, $h \in \Space{R}{} \setminus \{ 0 \}$, is
the subset of functions on $\Heisn$ defined by
\begin{equation} \label{hh}
\hilbh = \left\{ e^{2 \pi ihs} f (x,y) : E^j_h f = 0 \hspace{1cm} 1 \leq j \leq n \right\}
\end{equation}
where the operator $E^j_h = \frac{h}{2} (y_j + i c_i x_j)I + 2 \pi
(c_i \Partial{y_j}+ i \Partial{x_j})$ (this is the Fourier transform
of $D_h^j$ \eqref{eq:Cauchy-Riemann}). The inner product on $\hilbh$ is defined as 

\begin{equation} \label{hhip} \langle v_1 , v_2 \rangle_{\hilbh} =
\left( \frac{4}{h} \right)^{n} \int_{\Space{R}{2n}} v_1 (s,x,y)
\overline{v}_2 (s,x,y) \, dx \, dy.
\end{equation}
\end{defn}
Note in equation (\ref{hhip}) there is no integration over the $s$
variable since for any two functions $v_1 = e^{2 \pi ihs} f_1 (x,y)$
and $v_2 = e^{2 \pi ihs} f_2 (x,y)$ in $\hilbh$
\begin{equation} \nonumber
\langle v_1 , v_2 \rangle = \int_{\Space{R}{2n}} e^{2 \pi ihs} e^{- 2
\pi ihs} f_1 (x,y) \bar{f}_2 (x,y) \, dx \, dy = \int_{\Space{R}{2n}}
f_1 (x,y) \bar{f}_2 (x,y) \, dx \, dy 
\end{equation}
and hence there is no $s$-dependence. It is important to note that all
the $\hilbh$ are shift-invariant and thus invariant under convolutions. 
Since the Fourier transform intertwines multiplication and
differentiation we have 
\begin{equation}
\hilbh = \left\{ e^{2 \pi ihs} \fort_{(x,y)} (f(q,p)) : f \in F^2(\horb) \right\}.
\end{equation}
$\hilbh$ is mapped into another Hilbert space $\iilbh$ by the Fourier
transform. This Hilbert space $\iilbh$ is 
\begin{equation} \nonumber
\iilbh = \left\{ j(h',q,p) = \delta(h'-h) f(q,p) : f \in \fock \right\},
\end{equation}
where $\delta$ is the Dirac delta distribution. The inner product for
$j_1(h',q,p) = \delta(h'-h) f_1(q,p)$ and  $j_2(h',q,p) = \delta(h'-h)
f_2(q,p)$ in $\iilbh$ is 
\begin{equation} \nonumber
\langle j_1 , j_2 \rangle_{\iilbh} = \left( \frac{4}{h} \right)^n
\int_{\Space{R}{2n+1}} j_1 (h',q,p) \overline{ j_2 (h',q,p) } \, dh'
\, dq  \, dp = \langle f_1 , f_2 \rangle_{\fock}. 
\end{equation}
We define a set of states for each $h \neq 0$ using $\hilbh$ (later in
this subsection we will define a set of states for $h\neq0$ which are
defined using a kernel and a set of states for $h=0$ by a kernel). 
\begin{defn} \textup{\cite{Brodlie03a}}
\label{de:h-states}
A \emph{\(\myhbar\)-state} corresponding to a vector $v\in \hilbh$
is defined on a \(p\)-mechanical observable  $B$ by
\begin{equation} \nonumber
\langle B * v , v \rangle_{\hilbh}.
\end{equation}
\end{defn}
For any vector $f \in \fock$ equation (\ref{stateinfock}) gives us a
corresponding state. We now introduce a map $\statem$ which maps
vectors in $\fock$ to vectors in $\hilbh$ 
\begin{equation} \label{eq:statemapforhnonzero}
\statem (f(q,p)) = e^{2 \pi ihs} \hat{f}(x,y).
\end{equation}
The following Theorem proves that the states corresponding to vectors
$f$ and $\statem f$ give the same expectation values for observables
$B$ and $\rho_h (B)$ respectively. 

\begin{thm} \textup{\cite{Brodlie03a}}
\label{hilbh-f2h}
For any observable $B$ and any $v_1, v_2 \in \hilbh$, $h \in
\Space{R}{} \setminus \{ 0 \}$, of the form \(v_1 (s,x,y) = \statem
f_1\), \(v_2 (s,x,y) = \statem f_2\)
 we have the relationship
\begin{equation} \label{eq:statesrelation}
\langle B * v_1 , v_2 \rangle_{\hilbh} = \langle \rho_{h} (B) f_1, f_2 \rangle_{\fock}.
\end{equation}
\end{thm}

\begin{proof}
From the Plancherel identity for \(\Space{R}{2n}\)  we have
\begin{equation} \label{eq:relhhih}
\langle B * v_1,v_2 \rangle_{\hilbh} = \langle \widehat{B * v_1}, \hat{v}_2 \rangle_{\iilbh}
\end{equation}
where again $\hspace{0.3cm}  \hat{} \hspace{0.3cm}$  is the Fourier
transform on the Heisenberg group as described in equation
(\ref{eq:fourier-transform}). Using (\ref{eq:ftandshift}) equation
(\ref{eq:relhhih}) can be written as 
\begin{equation} \label{eq:secondrelhhih}
\langle B * v_1,v_2 \rangle_{\hilbh} = \langle c_\myhbar^{n+1}\widehat{\int B (g) \lambda_l
(g)\, dg\, v_1} , \hat{v}_2 \rangle_{\iilbh}. 
\end{equation}
Using (\ref{eq:ftandshift}) equation (\ref{eq:secondrelhhih}) becomes
\begin{eqnarray} \nonumber
\langle B * v_1 , v_2 \rangle_{\hilbh} 
&=& \langle c_\myhbar^{n+1}\widehat{\int B (g) \rho_h (g)\, dg\,
v_1} , \hat{v}_2 \rangle_{\hilbh} \\ \nonumber 
&=& \left(\frac{4}{\myhbar}\right)^n \int \rho_h (B) \delta (h'-h) f_1(q,p) \overline{\delta (h'-h)
f_2(q,p)} \, dq \, dp \, dh' \\ \nonumber 
&=& \left(\frac{4}{\myhbar}\right)^n \int \rho_h (B) f_1(q,p) \overline{f_2} (q,p) \, dq \, dp.
\end{eqnarray}
Hence the result has been proved.
\end{proof}
Taking $v_1=v_2$ in (\ref{eq:statesrelation}) shows that the states
corresponding to $f$ and $\statem f$ will give the same expectation
values for $\rho_h (B)$ and $B$ respectively. If we take $B$ to be a
time development operator we can get probability amplitudes between
states $v_1 \neq v_2$. The map $\statem$ can be realised as a map from
the set of functionals on the quantum observables to the set of
functionals on the set of \(p\)-mechanical observables. This map is
the adjoint of $\rho_h$ when realised as a map from \(p\)-observables
to quantum observables. 

We now go on to show that each of these states can also be realised by
an appropriate kernel.
\begin{thm} \textup{\cite{Brodlie03a}}
If $l(s,x,y)$ is defined to be the kernel
\begin{equation} \label{eq:relationbetweenkernelandvector}
l(s,x,y) = \left(\frac{4}{h}\right)^n \int_{\Space{R}{2n}}
v((s,x,y)^{-1} (s',x',y')) \overline{v((s',x',y'))} \, dx' \, dy'. 
\end{equation}
then
\begin{equation} \nonumber
\langle B*v,v \rangle_{\hilbh} = \int_{\Heisn} B(s,x,y) l(s,x,y) \, ds \, dx \, dy.
\end{equation}
\end{thm}
\begin{proof}
It is easily seen that
\begin{eqnarray} \nonumber
\lefteqn{\langle B*v,v \rangle = \left(\frac{4}{h}\right)^n \int_{\Space{R}{2n}} \int_{\Heisn}
B((s,x,y)) v((s,x,y)^{-1} (s',x',y'))} \\ \nonumber 
&& \qquad \qquad \qquad \qquad \times \overline{v((s',x',y'))} \, ds
\, dx \, dy \, dx' \, dy' \\ \label{eq:vandlsxy} 
&=& \left(\frac{4}{h}\right)^n \int_{\Heisn} B((s,x,y)) \\ \nonumber 
&& \qquad \qquad \times \left( \int_{\Space{R}{2n}} v((s,x,y)^{-1}
(s',x',y')  )\overline{v((s',x',y'))}  \,dx' \, dy' \right) \,ds \, dx
\, dy 
\end{eqnarray}
Note that there is no integration over $s'$ by the definition of the
$\hilbh$ inner product. 
\end{proof}

The quantum states defined through their kernels instead of vectors of a
Hilbert space are particularly suitable for contextual probability
interpretation~\cite{Khrennikov01a,Vaxjo01,Kisil01c} of quantum
mechanics, see the discussion in Section~\ref{sec:conclusions}. Thus we collect them
together under the following definition.
\begin{defn} \textup{\cite{Brodlie03a}}
We denote the set of kernels corresponding to the
elements in $\hilbh$ as $\lkerh$.
\end{defn}

Now we introduce $(q,p)$-states in  \(p\)-mechanics, which correspond to
classical states, they are again functionals on the \(C^*\)-algebra of
\(p\)-mechanical observables. Pure states in classical mechanics
evaluate observables at particular points of phase space, they can be
realised as kernels $\delta(q-q',p-p')$ for fixed $q$, $p$ in phase space,
that is
\begin{equation} \label{eq:classicalpurestateeval}
\int_{\Space{R}{2n}} F(q,p) \delta(q-q',p-p') \, dq' \, dp' = F(q,p).
\end{equation}
We now give the \(p\)-mechanical equivalent of pure classical states.
\begin{defn} \textup{\cite{Brodlie03a}}
\label{th:q-p-states}
A $(q,p)$-pure state is defined to be the set of
functionals, $k_{(0,q,p)}$, for fixed $(q,p) \in \Space{R}{2n}$ which
act on observables by 
\begin{equation} \label{eq:pclasspurestates} k_{(0,q,p)}(B(s,x,y)) =
\int_{\Heisn} B(s,x,y) e^{-2\pi i (qx+py)}\, dx \, dy.
\end{equation}
Each classical pure state $k_{(0,q,p)}$ is defined entirely by its kernel $l_{(0,q,p)}$
\begin{equation}
l_{(0,q,p)} = e^{-2\pi i (qx+py)}.
\end{equation}
\end{defn}
By equation (\ref{eq:classical-observables-1}) we have
\begin{equation} \label{eq:pmechanicalpurestateeval}
\int_{\Heisn} B(s,x,y) e^{-2\pi i (qx+py)}\, ds \, dx \, dy = F(q,p)
\end{equation}
where $F$ is the classical observable corresponding to $B$ (using the
relation exhibited in subsection \ref{sec:weyl-quantisation}), hence
when we apply state $k_{(0,q,p)}$ to a \(p\)-mechanical observable we
get the value of its classical counterpart at the point $(q,p)$ of
phase space.  We introduce the map $\statemo$ which maps classical
pure state kernels to \(p\)-mechanical classical pure state kernels
\begin{equation} \nonumber
\statemo (\xi(q,p)) = \hat{\xi} (x,y).
\end{equation}
This equation is almost identical to the relation in equation
(\ref{eq:statemapforhnonzero}). The kernels $\labo = e^{-2\pi i
(qx+py)}$, are the Fourier transforms of the delta functions
$\delta(q-q',p-p')$, hence pure $(q,p)$ states are just the
image of pure classical states.

Mixed states, as used in statistical mechanics \cite{Honerkamp98}, are
linear combinations of pure states. In \(p\)-mechanics $(q,p)$ mixed
states are defined in the same way.
\begin{defn}  \textup{\cite{Brodlie03a}}
Define $\lkero$, to be the space of all linear combinations of $(q,p)$
pure state kernels $l_{(0,q,p)}$, that is the set of all kernels
corresponding to $(q,p)$ mixed states. 
\end{defn}
The map $\statemo$ exhibits the same relations on mixed states as pure
states due to the linearity of the Fourier transform. Note that if we
consider the map $\statemo$ as mapping from functional to functional,
that is going from the dual space of classical observables on $\oorb$
to the dual space of the set of \(p\)-mechanical observables then it
is the adjoint of $\rho_{(q,p)}$. 

\begin{rem}
  \label{re:classical-states-through-vectors}
  The above description of classical states corresponds to quantum
  states defined through their
  kernels~\eqref{eq:relationbetweenkernelandvector}. It is possible to
  define classical states through vector in the Hilbert space
  \(\FSpace{L}{2}(\orbit{0})\) as well. Indeed a classical observable
  \(B(q,p)\) acts on \(\FSpace{L}{2}(\orbit{0})\) by
  multiplication. Then a vector \(v(q,p)\in\FSpace{L}{2}(\orbit{0})\)
  defines the state by the natural formula
  \(B(q,p)\mapsto\scalar{B(q,p)v(q,p)}{v(q,p)}\). This permits
  quantum superpositions of states while the dynamics of observables is
  governed by the classical Hamilton equation~\eqref{eq:Hamilton-equation}.
\end{rem}

In accordance with the general theory of \(C^*\)-algebras mentioned in
the beginning  of this subsection we could now describe a general
\(p\)-mechanical state:

\begin{prop}
  An arbitrary \(p\)-mechanical state is a superposition of quantum
  \(\myhbar\)-states given by Definition~\ref{de:h-states} and
  classical \((q,p)\)-states described in
  Definition~\ref{th:q-p-states}. 
\end{prop}

Consequently a comprehensive study of \(p\)-mechanical states, notably
their dynamics, could be done through this decomposition. The various
relations between \(p\), \(\myhbar\), and \((q,p)\)-states could be derived
from Figure~\ref{fi:quant-class-p-observ}. Indeed since all types of
states form the dual spaces to the corresponding spaces of
observables, the reversion of arrows on
Figure~\ref{fi:quant-class-p-observ} provides the maps between states
through the adjoint operators to \(\uir{\myhbar}\), \(\uir{0}\),
\(\oper{W}_r\), \(\oper{W}_0\), \(\oper{E}\), \(\oper{Q}_\myhbar\),
\(\oper{C}_{\myhbar\rightarrow 0}\). The adjoint operator to the wavelet
transform \(\oper{W}_r\) was identified with the \emph{inverse wavelet
  transform} in~\cite{Kisil98a}.


We conclude this section by the following result describing relations
between eigenvectors in \(\FSpace {F}{2}\) and their images under
\(\statem\)

\begin{thm}  \textup{\cite{Brodlie03a}}
For a \(p\)-observable $B \in \loneh$ and $f_1 \in \fock$, $\rho_h (B) f_1
= \lambda f_1$, if and only if for $v_1 (s,x,y) = \statem f_1 = e^{2
\pi ihs} \hat{f_1} (x,y) \in \hilbh$
\begin{equation} \nonumber
\langle B * v_1, v_2 \rangle = \lambda \langle v_1, v_2 \rangle
\end{equation}
holds for all $v_2 \in \hilbh$.
\end{thm}


\section{$p$-Mechanics: Dynamics}
\label{sec:p-mechanics-dynamics}
We introduce the \(p\)-mechanical brackets which fulfil all essential
physical requirements and have a non-trivial classical representation
coinciding with the Poisson brackets. A consistent \(p\)-mechanical
dynamic equation for observables is given in
subsection~\ref{sec:p-dynamic-equation}. In subsection
\ref{subsect:timeevolofstates} we give two equivalent dynamic
equations for \(p\)-mechanical states. Symplectic automorphisms of the
Heisenberg groups produce symplectic symmetries of \(p\)-mechanical,
quantum, and classical dynamics in
subsection~\ref{sec:quant-from-sypl}.

\subsection{$p$-Mechanical Brackets and  Dynamic Equation on
  $\Space{H}{n}$} 
\label{sec:p-dynamic-equation}

\label{sec:p-mechanical-bracket}

Having observables as convolutions on \(\Space{H}{n}\) we need a
dynamic equation for their evolution. To this end we seek a time
derivative generated by the observable associated with energy. The
first candidate is the derivation coming from
commutator~\eqref{eq:commutator}. However the straight commutator
  has at least two failures. The first failure is that it can't
  produce any dynamics on \(\orbit{0}\)~\eqref{eq:orbit-0}, see
  Remark~\ref{re:triv-commutator}.
The second failure is due to a mismatch in units: the \(p\)-mechanical
energy, $B_H$, is measured in units $ML^2/T^2$ whereas the time
derivative should be measured in $1/T$, that is the mismatch is in
units of action $ML^2/T$.

Fortunately, there is a possibility to fix both the above defects of
the straight commutator at the same time.  Let us define a multiple
\(\anti\) of a right inverse operator to the vector field
\(S\)~\eqref{eq:h-lie-algebra} on \(\Space{H}{n}\) by its actions on
exponents---characters of the centre \(Z\in\Space{H}{n}\):
\begin{equation}
  \label{eq:def-anti}
 S\anti=4\pi^2 I, \qquad \textrm{ where }\quad
  \anti e^{ 2\pi\rmi \myhbar s}=\left\{
    \begin{array}{ll}
      \displaystyle \frac{2\pi}{\rmi\myhbar\strut} e^{2\pi\rmi\myhbar
      s}, & \textrm{if } \myhbar\neq 0,\\ 4\pi^2 s\strut, & \textrm{if
      } \myhbar=0.
    \end{array} \right.
\end{equation} An alternative definition of \(\anti\) as a convolution
with a distribution is given in~\cite{Kisil00a}.

We can extend \(\anti\) by linearity to the entire space
\(\FSpace{L}{1}(\Space{H}{n})\). As a multiple of a right inverse to
\(S\) the operator \(\anti\) is measured in \(T/(ML^2)\)---exactly
that we need to correct the mismatch of units in the straight
commutator.  Thus we introduce~\cite{Kisil00a} a modified convolution
operation \(\star\) on \(\FSpace{L}{1}(\Space{H}{n})\):
\begin{equation}
    \label{eq:u-star} B'\star B = (B'* B)\anti
\end{equation} and the associated modified commutator
(\(p\)-mechanical brackets):
\begin{equation}
  \label{eq:u-brackets}
    \ub{B'}{B}=[B',B]\anti=B'\star B-B\star B'.
\end{equation} Obviously~\eqref{eq:u-brackets} is a bilinear
antisymmetric form on the convolution kernels. It was also
demonstrated in~\cite{Kisil00a} that the \(p\)-mechanical brackets
satisfy the Leibniz and Jacoby identities. They are all important for
consistent dynamics~\cite{CaroSalcedo99} along with the dimensionality
condition given in the beginning of this subsection.

From~\eqref{eq:rho-extended-to-L1} one gets \(\uir{\myhbar}(\anti
B)=\frac{2\pi}{i\myhbar}\uir{\myhbar}(B)\) for \(\myhbar\neq
0\). Consequently the modification of the commutator for
\(\myhbar\neq0\) is only slightly different from the original one:
\begin{equation}
  \label{eq:rho-h-repres-of-p-bracket} \uir{\myhbar}\ub{B'}{B} =
  \frac{1}{\rmi\hbar}[\uir{\myhbar}(B'),\uir{\myhbar}(B)], \qquad
  \textrm{ where } \hbar=\frac{\myhbar}{2\pi}\neq0.
\end{equation} The integral representation of the modified commutator
kernel becomes (cf.~\eqref{eq:repres-commutator}):
\begin{equation}
  \fl
  \label{eq:repres-ubracket}
   \ub{B'}{B}\!\hat{_s}
    = c_\myhbar^{n}\int_{\Space{R}{2n}}\!
  \frac{4\pi}{\myhbar}\sin\left(\pi\myhbar (xy'-yx')\right)
  \hat{B}'_s(\myhbar ,x',y') \hat{B}_s(\myhbar ,x-x',y-y') \,dx'dy',
\end{equation} where we may understand the expression under the
integral as
\begin{equation}
  \label{eq:convolution-kernel}
  \frac{4\pi}{\myhbar}\sin\left({\pi\myhbar} (xy'-yx')\right)
  =4\pi^2\sum_{k=1}^\infty (-1)^{k+1} \left(\pi\myhbar\right)^{2(k-1)}
  \frac{ (xy'-yx')^{2k-1}}{(2k-1)!}
\end{equation} This makes the operation~\eqref{eq:repres-ubracket} for
\(\myhbar=0\) significantly distinct from the vanishing
integral~\eqref{eq:repres-commutator}. Indeed it is natural to assign
the value \(4\pi^2(xy'-yx')\) to~\eqref{eq:convolution-kernel} for
\(\myhbar=0\). Then the integral in~\eqref{eq:repres-ubracket} becomes
the Poisson brackets for the Fourier transforms of \(B'\) and \(B\)
defined on \(\orbit{0}\)~\eqref{eq:orbit-0}:
\begin{equation}
  \label{eq:Poisson}
    \uir{(q,p)}\ub{B'}{B} = \frac{\partial \hat{B}'(0,q,p)}{\partial
    q} \frac{\partial \hat{B}(0,q,p)}{\partial p} -\frac{\partial
    \hat{B}'(0,q,p)}{\partial p} \frac{\partial
    \hat{B}(0,q,p)}{\partial
      q}.
\end{equation} The same formula is obtained~\cite[Prop.~3.5]{Kisil00a}
if we directly calculate \(\uir{(q,p)}\ub{B'}{B}\) rather than resolve
the indeterminacy for \(\myhbar= 0\)
in~\eqref{eq:convolution-kernel}. This means there is continuity in
our construction at \(\myhbar=0\) which represents the
\emph{correspondence principle} between quantum and classical
mechanics.

We have now arrived at the conclusion the \emph{Poisson brackets
  and the inverse of the Planck constant \(1/\myhbar\) have the same
  dimensionality because they are the image of the same object
  (anti-derivative~\eqref{eq:def-anti}) under different
  representations \eqref{eq:stone-inf} and \eqref{eq:stone-one} of the
  Heisenberg group.}

Note that functions \(X=\delta(s)\delta^{(1)}(x)\delta(y)\) and
\(Y=\delta(s)\delta(x)\delta^{(1)}(y)\) (see~\eqref{eq:p-mech-q}
and~\eqref{eq:p-mech-p}) on \(\Space{H}{n}\) are measured in units
\(L\) and \(ML/T\) (inverse to \(x\) and \(y\)) correspondingly as
respective derivatives of the dimensionless function
\(\delta(s)\delta(x)\delta(y)\). Then the \(p\)-mechanical brackets
\(\ub{X}{\cdot}\) and \(\ub{Y}{\cdot}\) with those functions have
dimensionality of \(T/(ML)\) and \(1/L\) correspondingly. Their
representation \(\uir{*} \ub{X}{\cdot}\) and \(\uir{*} \ub{Y}{\cdot}\)
(for both type of representations \(\uir{\myhbar}\) and
\(\uir{(q,p)}\)) are measured by \(L\) and \(ML^2/T\) and are just
derivatives:
\begin{equation}
  \label{eq:shifts-on-orbits} \uir{*} \ub{X}{\cdot} = \frac{\partial
  }{\partial p}, \qquad \uir{*} \ub{Y}{\cdot} = \frac{\partial
  }{\partial q} .
\end{equation} Thus \(\uir{*} \ub{X}{\cdot}\) and \(\uir{*}
\ub{Y}{\cdot}\) are generators of shifts on both types of orbits
\(\orbit{\myhbar}\) and \(\orbit{0}\) independent from \(\myhbar\).

Since the modified commutator~\eqref{eq:u-brackets} with a
\(p\)-mechanical energy has the dimensionality \(1/T\)---the same as
the time derivative---we introduce the dynamic equation for an
observable \(B(s,x,y)\) on \(\Space{H}{n}\) based on that modified
commutator as follows
\begin{equation}
  \fl
  \label{eq:p-equation}
  \frac{d B}{d t}=\ub{B}{B_H} .
\end{equation} \begin{rem}
  It is a general tendency to make a Poisson bracket or quantum
  commutator out of any two observables and say that they form a Lie
  algebra. However there is a physical meaning to do that if at least
  one of the two observables is an energy, coordinate or momentum: in
  these cases the bracket produces the time
  derivative~\eqref{eq:p-equation} or corresponding shift
  generators~\eqref{eq:shifts-on-orbits} of the other observable.
\end{rem} A simple consequence of the previous consideration is that
the \(p\)-dynamic equation~\eqref{eq:p-equation} is reduced
\begin{enumerate}
\item by the representation \(\uir{\myhbar}\),
      \(\myhbar\neq0\)~\eqref{eq:stone-inf} on
        \(\FSpace{F}{2}(\orbit{\myhbar})\)~\eqref{eq:co-adjoint-orbits-inf}
        to Moyal's form of Heisenberg equation \cite[(8)]{Zachos02a}
        based on the formulae~\eqref{eq:rho-h-repres-of-p-bracket}
        and~\eqref{eq:repres-ubracket}:
        \begin{equation}
          \label{eq:moyal-equation} \frac{d \uir{\myhbar}(B)}{d t}
          =\frac{1}{i\hbar}[\uir{\myhbar}(B), H_\myhbar ], \qquad
          \textrm{
            where the operator } H_\myhbar =\uir{\myhbar}(B_H);
        \end{equation}
\item by the representations \(\uir{(q,p)}\)~\eqref{eq:stone-one} on
        \(\displaystyle\orbit{0}\)~\eqref{eq:orbit-0} to Poisson's
        equation \cite[\S~39]{Arnold91} based on the
        formula~\eqref{eq:Poisson}:
        \begin{equation}
          \label{eq:Hamilton-equation} \frac{d \hat{B}}{d
          t}=\{\hat{B}, H\} \qquad \textrm{ where the function }
          H(q,p)=\uir{(q,p)}(B_H)=\hat{B_H}\left(0,{q},{p}\right).
        \end{equation}
\end{enumerate} The same connections are true for the solutions of the
three equations~\eqref{eq:p-equation}--\eqref{eq:Hamilton-equation},
this equation is demonstrated in section \ref{sect:examplesjointp}.

\subsection{$p$-Mechanical Dynamics for States}
\label{subsect:timeevolofstates} We now go on to show how
\(p\)-mechanical states evolve with time.  We first show how the
elements of $\lkerh$, for all $h\in\Space{R}{}$ evolve with time and
that this time evolution agrees with the time evolution of
\(p\)-observables. In doing this we show that for the particular case of
$\lkero$ the time evolution is the same as classical states under the
Liouville equation. Then we show how the elements of $\hilbh$ evolve
with time and prove that they agree with the Schr\"odinger picture of
motion in quantum mechanics. Before we can do any of this we need to
give the definition of a Hermitian convolution.

\begin{defn} 
\textup{\cite{Brodlie03a}}
We call a \(p\)-mechanical observable \(B\) Hermitian if it
corresponds to a Hermitian convolution, that is for any functions
$f_1,f_2$ on the Heisenberg group
\begin{equation} \nonumber
\int_{\Heisn} (B*f_1)(g) \overline{f_2(g)} dg = \int_{\Heisn} f_1(g)\overline{(B*f_2)(g)} dg. 
\end{equation}
\end{defn}
If a \(p\)-observable $B$ is Hermitian then  $ B(g) =
\overline{B(g^{-1})}$, this is the result of a trivial
calculation. From now on we denote $\overline {B (g^{-1})}$ as
$B^*$. For our purposes we just need to assume that the distribution
or function, $B$, corresponding to the observable is real and
$B(s,x,y)=B(-s,-x,-y)$. 

\begin{defn} 
\textup{\cite{Brodlie03a}}
If we have a system with energy $B_H$ then an arbitary
kernel $l \in \lkerh$, $h\in \Space{R}{}$, evolves under the equation
\begin{equation} \label{eq:zerostatestimeeq}
\Fracdiffl{l}{t} =  \ub{B_H}{l}{}.
\end{equation}
\end{defn}
We now show that the time evolution of these kernels coincides with the
time evolution of \(p\)-mechanical observables. 
\begin{thm} 
\label{thm:timeevolofkernelisok}
\textup{\cite{Brodlie03a}}
If $l$ is a kernel evolving under equation (\ref{eq:zerostatestimeeq})
then for any observable $B$ 
\begin{equation} \nonumber
\Diffl{t} \int_{\Heisn} B \, l \, dg = \int_{\Heisn} \ub{B}{B_H}{} \, l \, dg.
\end{equation}
\end{thm}
\begin{proof}
This result can be verified by the direct calculation,
\begin{eqnarray} \nonumber
\lefteqn{\Diffl{t} \int_{\Heisn} B(s,x,y) l(s,x,y) \, ds \, dx \, dy } \\ \nonumber
&=& \int_{\Heisn} B(s,x,y) \antid (B_H *l - l *B_H ) (s,x,y) \, ds \, dx \, dy \\ \label{eq:usedintbypartsintime}
&=& - \int_{\Heisn} \antid B(s,x,y) (B_H * l - l *B_H ) (s,x,y) \, ds \, dx \, dy \\ \nonumber
&=& \int_{\Heisn} \antid ((B * B_H)(s,x,y) l (s,x,y) \\  \label{eq:usedhermitianinzero}
&& \qquad -(B_H * B)(s,x,y) l (s,x,y) ) \, ds \, dx \, dy \\ \nonumber
&=& \int_{\Heisn} \ub{B}{B_H}{}(s,x,y) l(s,x,y) \, ds \, dx \, dy.
\end{eqnarray}
At (\ref{eq:usedintbypartsintime}) we have used integration by parts
while (\ref{eq:usedhermitianinzero}) follows since $B_H$ is
Hermitian. 
\end{proof}
If we take the representation $\rho_{(q,p)}$ of equation
(\ref{eq:zerostatestimeeq}) we get the Liouville equation
\cite[Eq. 5.42]{Honerkamp98} for a kernel $\statemo^{-1}(l)$ moving in
a system with energy $\rho_{(q,p)} (B_H)$. This only holds for
elements in $\lkero$ and can be verified by a similar calculation to
\cite[Propn. 3.5]{Kisil00a}. 

We now show how the vectors in $\hilbh$ evolve with time. First we
extend our definition of $\antid$ which was initially introduced in
equation (\ref{eq:def-anti}). $\antid$ can also be defined as an
operator on each $\hilbh$, $h \in \Space{R}{} \setminus \{ 0 \}$,
$\antid:\hilbh \mapsto \hilbh$ by
\begin{equation} \nonumber
\antid v = \frac{2\pi}{ih} v.
\end{equation}
As the derivative operator the antiderivative $\antid$ is
skew-symmetric, i.e. $\antid^*=-\antid$, on each $\hilbh$, $h \in
\Space{R}{} \setminus \{ 0 \}$. 
\begin{defn}
\textup{\cite{Brodlie03a}}
If we have a system with energy $B_H$ then an arbitrary vector $v \in
\hilbh$ evolves under the equation 
\begin{equation} \label{eq:timevolinhh}
\Fracdiffl{v}{t} = \antid B_H *v = B_H * \antid v
\end{equation}
\end{defn}
The operation of left convolution preserves each $\hilbh$ so this time
evolution is well defined. Equation (\ref{eq:timevolinhh}) implies
that if we have $B_H$ time-independent then for any $v \in \hilbh$ 
\begin{equation} \nonumber
v(t;s,x,y) = e^{t  \antid B_H } v(0;s,x,y)
\end{equation}
where $e^{ \antid B_H } $ is the exponential of the operator of
applying the left convolution of $B_H$ and then applying
$\antid$. There is no mismatch in units here since $\antid$ has units
$T/ML^2$ and $B_H$ has units $ML^2/T^2$, hence their product has units
$1/T$. 

\begin{thm} 
\label{thm:schroheis} 
\textup{\cite{Brodlie03a}}
If we have a system with energy
$B_H$ (assumed to be Hermitian) then for any state $v \in \hilbh$ and
any observable $B$

\begin{equation} \nonumber \frac{d}{dt} \langle B * v, v \rangle =
\langle \ub{B}{B_H}{} * v , v \rangle.
\end{equation}
\end{thm}

\begin{proof} The result follows from the direct calculation:
\begin{eqnarray} \nonumber
\frac{d}{dt} \langle B * v (t) , v (t) \rangle
&=& \langle B * \frac{d}{dt} v , v \rangle + \langle B * v,
\frac{d}{dt} v \rangle \\ \nonumber 
&=& \langle B * \antid B_H *v  , v \rangle + \langle B * v, \antid B_H
*v \rangle  \\ \label{eq:usedadjofantid} 
&=& \langle B* \antid B_H * v, v \rangle - \langle \antid B * v , B_H
*v \rangle \\ \label{eq:usedhermitian} 
&=& \langle B * \antid B_H * v  , v \rangle - \langle \antid  B_H * B
* v, v \rangle  \\ \nonumber 
&=& \langle \ub{B}{B_H}{} *v , v \rangle. 
\end{eqnarray}
Equation (\ref{eq:usedadjofantid}) follows since $\antid$ is
skew-adjoint. At (\ref{eq:usedhermitian}) we have used the fact that
$B_H$ is Hermitian. 
\end{proof}
This Theorem tells us that the time evolution of states in $\hilbh$
coincides with the time evolution of observables as described in
equation (\ref{eq:p-equation}). We now give a Corollary to show that
the time evolution of \(p\)-mechanical states in $\hilbh$, $h \in
\Space{R}{} \setminus \{ 0 \} $ is the same as the time evolution of
quantum states. 
\begin{cor}
\textup{\cite{Brodlie03a}}
If we have a system with energy $B_H$ (assumed to be Hermitian) and an
arbitrary state $v = \statem f = e^{2 \pi ihs} \hat{f} (x,y)$
(assuming $h \neq 0$) then for any observable $B (t;s,x,y)$ 
\begin{equation} \nonumber
\Diffl{t} \langle B * v (t) , v (t) \rangle_{\hilbh} = \Diffl{t} \langle \rho_h (B) f(t) , f(t) \rangle_{\fock}.
\end{equation}
Where $\Fracdiffl{f}{t} = \frac{1}{ih} \rho_{h} (B_H) f$ (this is just
the usual Schr\"odinger equation). 
\end{cor}
\begin{proof}
From Theorem \ref{thm:schroheis} we have
\begin{eqnarray} \nonumber
\frac{d}{dt} \langle  B * v, v \rangle &=&  \langle \ub{B}{B_H}{} * v , v \rangle \\ \nonumber
&=& \langle \antid (B * B_H - B_H * B ) * v, v \rangle \\ \nonumber
&=& \langle (B * B_H - B_H * B ) * \antid v, v \rangle \\ \nonumber
&=& \frac{2 \pi}{ih} \langle (B * B_H - B_H * B ) * v, v \rangle \\ \nonumber
&=& \frac{1}{i \hbar} ( \langle B * B_H *v,v \rangle - \langle B * v, B_H * v \rangle)
\end{eqnarray}
The last step follows since $B_H$ is Hermitian. Using equation
(\ref{eq:statesrelation}), the above equation becomes, 
\begin{eqnarray} \nonumber
\frac{d}{dt} \langle B * v, v \rangle &=& \frac{1}{i\hbar} (\langle \rho_h (B) \rho_h (B_H)f,f \rangle_{\fock} - \langle \rho_h (B) f , \rho_h (B_H )f \rangle_{\fock} )\\ \nonumber
&=& \frac{d}{dt} \langle \rho_h (B)f, f \rangle_{\fock},
\end{eqnarray}
which completes the proof.
\end{proof}

Hence the time development in $\hilbh$ for $h \neq 0$ gives the same
time development as in $\fock$.
If $l(s,x,y) = \left( \frac{4}{h} \right)^n \int_{\Heisn}
\overline{v((s',x',y'))} v((s',x',y')^{-1} (s,x,y)) \, dx' \, dy'$
then by Theorems \ref{thm:timeevolofkernelisok} and
\ref{thm:schroheis} we have that
\begin{equation} \label{eq:sametimeevolofkernelandvector}
\Diffl{t} \langle B*v,v \rangle_{\hilbh} = \Diffl{t} \int_{\Heisn} B \, l \, dg.
\end{equation}

We conclude this subsection touching the question on \emph{mixing}
between quantum and classical states.  A simple application of the
representation theory yields the following ``no-go'' results equivalent to
the main conclusion of the paper~\cite{Salcedo96}:
\begin{thm}
  \label{th:no-q-c-mixing}
  If the Hamiltonian of a \(p\)-mechanical system is given by a
  convolution operator then there is no mixing between quantum and
  classical states during the induced evolution.
\end{thm}
Obviously, this results essentially relies on the assumption that
the Hamiltonian is a convolution operator. Examples of mixing for
quantum and classic states for more general Hamiltonians will be
discussed somewhere else.

\subsection{The $p$-Mechanical Interaction Picture}
\label{sect:interactpic}

In the Schr\"odinger picture, time evolution is governed by the states
and their equations $\frac{d v}{dt} = \antid B_H * v$
$\Fracdiffl{l}{t}= \ub{B_H}{l}{}$. In the Heisenberg picture, time
evolution is governed by the observables and the equation $\frac{d
B}{dt} = \ub{B}{B_H}{}$. In the interaction picture we divide the time
dependence between the states and the observables. This is suitable
for systems with a Hamiltonian of the form $B_H = B_{H_0} + B_{H_1}$
where $B_{H_0}$ is time independent. The interaction picture has many
uses in perturbation theory \cite{Kurunoglu62}.

Let a \(p\)-mechanical system have the Hamiltonian $B_H = B_{H_0} +
B_{H_1}$ where $B_{H_0}$ is time independent. We first describe the
interaction picture for elements of $\hilbh$. Define $\exp (t \antid
B_{H_0})$ as the operator on $\hilbh$ which is the exponential of the
operator of convolution by $t \antid B_{H_0}$. Now if $B$ is an
observable let
\begin{equation} \label{eq:obsevolveininter}
\tilde{B} = \exp(t \antid B_{H_0}) B \exp(-t \antid B_{H_0})
\end{equation}
If $v \in \hilbh$, define $\tilde{v} = (\exp (- t \antid B_{H_0} )) v$, then we get
\begin{eqnarray} \label{eq:gentimeevolofinter}
\frac{d}{dt} \tilde{v} &=& \frac{d}{dt} ( \exp(-t \antid B_{H_0}) v ) \\ \nonumber
&=& - \antid B_{H_0} * \tilde{v} + \exp(-t \antid B_{H_0}) ( \antid (B_{H_0} + B_{H_1}) *v) \\ \nonumber
&=& - \antid B_{H_0} * \tilde{v} + \antid B_{H_0} * \exp(-t \antid B_{H_0}) v + \exp(-t \antid B_{H_0}) \antid B_{H_1} v \\ \nonumber
&=& (\exp(-t \antid B_{H_0}) \antid B_{H_1} \exp(t \antid B_{H_0}))(\tilde{v})
\end{eqnarray}
Now we describe the interaction picture for a state defined by a kernel $l$. Define
\begin{equation} \nonumber
\tilde{l} = e^{-t\ub{B_{H_0}}{\cdot}{}} l =  \exp(-t \antid B_{H_0}) l \exp(+t \antid B_{H_0})
\end{equation}
then
\begin{eqnarray} \nonumber
\Fracdiffl{\tilde{l}}{t} &=& \antid B_{H_0} * \tilde{l} + \exp(-t \antid B_{H_0}) \ub{B_{H_0} + B_{H_1}}{l}{} \exp(t \antid B_{H_0}) - \tilde{l} * \antid B_{H_0} \\ \nonumber
&=& \exp(-t \antid B_{H_0}) \ub{B_{H_1}}{l}{} \exp(t \antid B_{H_0}) \\ \nonumber
&=& \exp(-t \antid B_{H_0}) (\antid (B_{H_1} * \exp(t \antid B_{H_0}) \tilde{l} \exp(-t \antid B_{H_0}) \\ \nonumber
&& \qquad - \exp(t \antid B_{H_0}) \tilde{l} \exp(-t \antid B_{H_0}) * B_{H_1} )) \exp(t \antid B_{H_0}) \\ \label{eq:interforkernel}
&=& \ub{\exp(-t \antid B_{H_0}) B_{H_1} \exp(t \antid B_{H_0})} {\tilde{l}}{}
\end{eqnarray}
This shows us how interaction states evolve with time, while the
observables evolve by (\ref{eq:obsevolveininter}). Note that if we
take $B_{H_0}=B_H$ we have the Heisenberg picture, while if we take
$B_{H_1} = B_H$ we have the Schr\"odinger picture. The interaction
picture is very useful in studying the forced harmonic oscillator as
will be shown in subsection \ref{sect:interpictofforcedosc}. 

\subsection{Symplectic Invariance from Automorphisms of
$\Space{H}{n}$} \label{sec:quant-from-sypl}

\begin{figure}[tbp]
  \begin{center}
      \includegraphics{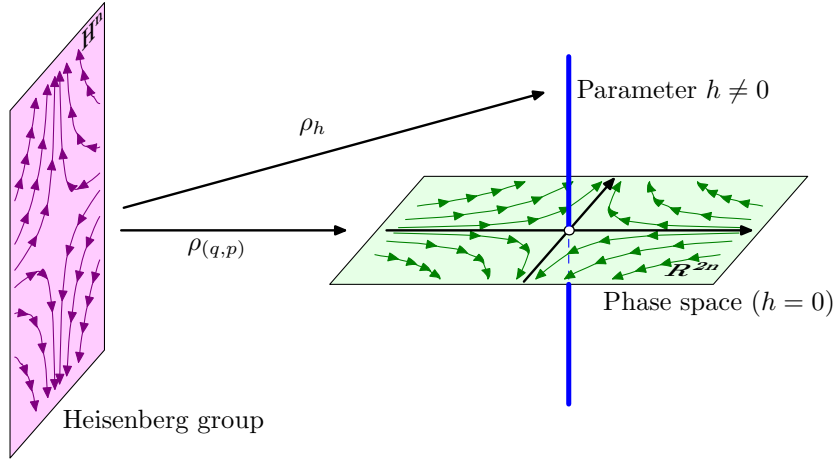}
  \end{center}
    \caption[Symplectic symmetries of quantum and classical
    mechanics]{
        \href{http://maths.leeds.ac.uk/~kisilv/nccg2.20.gif}{
          Automorphisms of
        \(\Space{H}{n}\) generated by the symplectic group \({Sp}(n)\)
        do not mix representations \(\uir{\myhbar}\) with different
        Planck constants \(\myhbar\) and act by the metaplectic
        representation inside each of them. In the contrast those
        automorphisms of \(\Space{H}{n}\) act transitively on the set
        of one-dimensional representations \(\uir{(q,p)}\) joining
        them into the tangent space of the classical phase space
        \(\Space{R}{2n}\)}.}
    \label{fig:p-mechanics}
\end{figure}

Let \(A: \Space{R}{2n} \rightarrow \Space{R}{2n}\) be a linear
\emph{symplectomorphism}~\cite[\S~41]{Arnold91},
\cite[\S~4.1]{Folland89}, i.e. a map defined by the \(2n\times 2n\)
matrix:
\begin{displaymath}
  A: \twovect{x}{y} \mapsto
  \matr{a}{b}{c}{d}\twovect{x}{y}=\twovect{ax+by}{cx+dy}
\end{displaymath} preserving the symplectic
form~\eqref{eq:symplectic-form}:
\begin{equation}
  \label{eq:symplectic-inavriance}
  \omega\left(A(x,y);A(x',y')\right)=\omega(x,y;x',y').
\end{equation} All such transformations form the symplectic group
\(Sp(n)\).
It follows from the identities~\eqref{eq:symplectic-inavriance}
and~\eqref{eq:H-n-group-law} that the linear transformation \(\alpha :
\Space{H}{n} \rightarrow \Space{H}{n} \) such that
\(\alpha(s,x,y)=(s,A(x,y))\) is an automorphism of \(\Space{H}{n}\).
 Let us also denote by
\(\tilde{\alpha}=\tilde{\alpha}_A\) a unitary transformation of
\(\FSpace{L}{2}(\Space{H}{n})\) in the form
\begin{displaymath}
  \tilde{\alpha}(f)(s,x,y)=\sqrt{\det A}f(s,A(x,y)),
\end{displaymath} which is well defined~\cite[\S~4.2]{Folland89} on
the double cover \(\widetilde{Sp}(n)\) of the group \(Sp(n)\).  The
correspondence \(A\mapsto \tilde{\alpha}_A\) is a linear unitary
representation of the symplectic group in
\(\FSpace{L}{2}(\Space{H}{n})\). One can also check the intertwining
property
\begin{equation}
  \label{eq:alpha-intertwine-reg-repres} \lambda_{l(r)}(g)\circ
  \tilde{\alpha}=\tilde{\alpha}\circ \lambda_{l(r)}(\alpha(g))
\end{equation} for the left (right) regular
representations~\eqref{eq:left-right-regular} of \(\Space{H}{n}\).

Because \(\alpha\) is an automorphism of \(\Space{H}{n}\) the map
\(\alpha^*: B(g)\mapsto B(\alpha(g))\) is an automorphism of the
convolution algebra \(\FSpace{L}{1}(\Space{H}{n})\) with the
multiplication~\(*\)~\eqref{eq:de-convolution}, i.e.
\({\alpha^*}(B_1)*{\alpha^*} (B_2)={\alpha^*}(B_1*B_2)\). Moreover
\({\alpha^*}\) commutes with the
antiderivative~\(\anti\)~\eqref{eq:def-anti}, thus \(\tilde{\alpha}\)
is an automorphism of \(\FSpace{L}{1}(\Space{H}{n})\) with the
modified multiplication~\(\star\)~\eqref{eq:u-star} as well, that is
\begin{equation} \label{eq:canbeusedforschroeqn}
 {\alpha^*}(B_1)
\star{\alpha^*}(B_2)={\alpha^*}(B_1\star B_2).
\end{equation}
 By the linearity we
can extend the intertwining
property~\eqref{eq:alpha-intertwine-reg-repres} to the convolution
operator \(K\) as follows:
\begin{equation}
 \alpha^*K \circ\tilde{\alpha}=\tilde{\alpha}\circ K.
 \label{eq:intertwining}
\end{equation}

Since \(\alpha\) is automorphism of \(\Space{H}{n}\) it fixes the unit
\(e\) of \(\Space{H}{n}\) and its differential
\(d\alpha:\algebra{h}^n\rightarrow \algebra{h}^n\) at \(e\) is given
by the same matrix as \(\alpha\) in the exponential
coordinates. Obviously \(d\alpha\) is an automorphism of the Lie
algebra \(\algebra{h}^n\).  By the duality between \(\algebra{h}^n\)
and \(\algebra{h}^*_n\) we obtain the adjoint map
\(d\alpha^*:\algebra{h}^*_n\rightarrow \algebra{h}^*_n\) defined by
the expression
\begin{equation}
  \label{eq:d-alpha-star} d\alpha^*: (\myhbar,q,p) \mapsto
  (\myhbar,A^t(q,p)),
\end{equation} where \(A^t\) is the transpose of \(A\).  Obviously
\(d\alpha^*\) preserves any orbit
\(\orbit{\myhbar}\)~\eqref{eq:co-adjoint-orbits-inf} and maps the
orbit \(\orbit{(q,p)}\)~\eqref{eq:co-adjoint-orbits-one} to
\(\orbit{A^t(q,p)}\).

Identity~\eqref{eq:d-alpha-star} indicates that both representations
\(\uir{\myhbar}\) and \((\uir{\myhbar}\circ
\alpha)(s,x,y)=\uir{\myhbar}(s,A(x,y))\) for \(\myhbar\neq0\)
correspond to the same orbit \(\orbit{\myhbar}\). Thus they should be
equivalent,
i.e. there is an intertwining operator \(U_A:
\FSpace{F}{2}(\orbit{\myhbar}) \rightarrow
\FSpace{F}{2}(\orbit{\myhbar})\) such that \(U_A^{-1}
\uir{\myhbar}U_A=\uir{\myhbar}\circ \alpha\). Then the correspondence
\(\sigma: A \mapsto U_A\) is a linear unitary representation of the
double cover \(\widetilde{Sp}(n)\) of the symplectic group called the
\emph{metaplectic representation}~\cite[\S~4.2]{Folland89}.
Thus we have
\begin{prop}
  \textup{\cite{Kisil02e}}
  \label{pr:symplectic} The \(p\)-mechanical brackets are invariant
  under the symplectic automorphisms of \(\Space{H}{n}\):
  \(\ub{\tilde{\alpha} B_1}{\tilde{\alpha}
    B_2}= \tilde{\alpha} \ub{B_1}{B_2}\). Consequently the dynamic
  equation~\eqref{eq:p-equation} has symplectic symmetries which are
  reduced
\begin{enumerate}
\item by \(\uir{\myhbar}\), \(\myhbar\neq0\) on
  \(\orbit{\myhbar}\)~\eqref{eq:co-adjoint-orbits-inf} to the
  metaplectic representation in quantum mechanics;
\item by \(\uir{(q,p)}\) on
  \(\displaystyle\orbit{0}\)~\eqref{eq:orbit-0} to the symplectic
  symmetries of classical mechanics~\textup{\cite[\S~38]{Arnold91}}.
\end{enumerate} \end{prop}


Combining intertwining properties of all three
components~\eqref{eq:Weyl-quantisation-decomposition} in the Weyl
quantisation we get
\begin{cor} 
  \textup{\cite{Kisil02e}}
 The Weyl quantisation \(\oper{Q}_\myhbar\)~\eqref{eq:weyl-calculus}
  is the \emph{intertwining operator} between classical and
  metaplectic representations.
\end{cor}

The two equations for the time evolution of states
(\ref{eq:zerostatestimeeq}), (\ref{eq:timevolinhh}) are both invariant
under the symplectic automorphisms of $\Heisn$. The invariance of
equation (\ref{eq:timevolinhh}) is a consequence of
(\ref{eq:canbeusedforschroeqn}) while the invariance of
(\ref{eq:zerostatestimeeq}) follows from the invariance of the
\(p\)-mechanical brackets.

\subsection{Coherent States} \label{sect:cstates}

The coherent states defined in section \ref{sec:fock-type-spaces} all
had a function supported at $(0,0)\in\Space{R}{2n}$ as their classical limit, rather
than being supported around different classical states $(q,p)$. In
this section we rectify this problem by introducing an overcomplete
system of vectors in $\hilbh$ through a representation of
$\Heisn$. The states which correspond to these vectors are an
overcomplete system of coherent states for each $h \neq 0$. We then
show that these vectors correspond to a system of kernels in $\lkerh$,
whose limit is the $(q,p)$ pure state kernels.

Initially we need to introduce a vacuum vector in $\hilbh$.  The vector
in $\fock$ corresponding to the ground state is (c.f. equation
(\ref{eq:vacuum}))
\begin{equation} \nonumber
f_0 (q,p) = \exp \left( -\frac{2\pi}{h} (\omega m q^2 + (\omega
  m)^{-1} p^2 ) \right), \hspace{1cm} h>0 . 
\end{equation}
The image of this under $\statem$ is
\begin{equation} \nonumber
e^{2 \pi ihs} \fort ( f_0 ) = e^{2 \pi ihs} \int_{\Space{R}{2n}}
e^{-\frac{2 \pi}{h} (m \omega q^2 + (m \omega )^{-1} p^2)} e^{-2 \pi
  i(qx+py)} \, dq \, dp. 
\end{equation}
Using the basic formula
\begin{equation} \label{eq:thewaveletformula}
\int_{\Space{R}{}} \exp(-a x^2 + b x + c) dx = \left( \frac{\pi}{a}
\right)^{\frac{1}{2}} \exp \left( \frac{b^2}{4a} + c \right), \textrm{
  where  } a > 0 
\end{equation}
we get
\begin{equation} \nonumber
\statem (f_0 ) =  e^{2 \pi ihs} \fort ( f_0 ) = \left( \frac{h}{2}
\right)^n  \exp \left(  2 \pi ihs - \frac{\pi h}{2} \left(
    \frac{x^2}{\omega m} + y^2 \omega m \right) \right), 
\end{equation}
which is the element of $\hilbh$ corresponding to the ground state.

\begin{defn} 
\textup{\cite{Brodlie03a}}
Define the vacuum vector in $\hilbh$ as
\begin{equation} \nonumber
\voo = \left( \frac{h}{2} \right)^n \exp \left( 2\pi i hs - \frac{\pi
    h}{2} \left( \frac{x^2}{\omega m} + y^2 \omega m \right) \right), 
\end{equation}
where $\omega$ and $m$ are constants representing frequency and mass respectively.
\end{defn}
Now we calculate the kernel, $\loo$, for the ground state by the
relationship (\ref{eq:relationbetweenkernelandvector}) between kernels
and vectors. 

\begin{eqnarray} \nonumber 
\lefteqn{\loo (s,x,y) }\nonumber\\
&=&
\left(\frac{4}{h} \right)^n \int_{\Space{R}{2n}} \voo
((-s,-x,-y)(s',x',y')) \overline{\voo(s',x',y')} \, dx' \, dy'  \nonumber\\
\label{eq:usedthewaveletformula} &=& \exp \left( -2\pi ihs -\frac{\pi
h}{2}(\frac{x^2}{\omega m} +\omega m y^2) \right) \\ \nonumber &&
\qquad \times \exp \left(\frac{\pi h}{4} \left( \omega m
\left(iy+\frac{x}{\omega m}\right)^2 + \frac{1}{\omega m} ( \omega m y
-ix)^2 \right) \right)
\end{eqnarray}
at (\ref{eq:usedthewaveletformula}) we have used formula
(\ref{eq:thewaveletformula}). By a simple calculation it can be shown
that 
\begin{equation} \nonumber
\omega m \left( iy+\frac{x}{\omega m}\right)^2 + \frac{1}{\omega m}
(\omega m y -ix)^2  = 0 
\end{equation}
hence
\begin{equation} \nonumber
\loo = \exp \left( -2\pi ihs -\frac{\pi h}{2}\left( \frac{x^2}{\omega
      m} +\omega m y^2 \right) \right). 
\end{equation}
Recalling functions $X$ and $Y$ from equations ~\eqref{eq:p-mech-q}
and~\eqref{eq:p-mech-p}
\begin{displaymath}
\begin{array}{ccc}
X= \frac{1}{2 \pi i} \zerodelxone & \textrm{ and } & Y = \frac{1}{2 \pi i} \zerodelyone.
\end{array}
\end{displaymath}
Under left and right convolution $X$ and $Y$ generate
left~\eqref{eq:der-repr-h-bar} and
right~\eqref{eq:der-repr-h-bar-right} 
invariant vector fields respectively. That is, if $B$ is a function or
distribution on $\Heisn$ then 
\begin{eqnarray*}
X*B = \frac{1}{2 \pi i}\left(\Partial{x} - \frac{y}{2}
  \Partial{s}\right)B, & \hspace{1cm} & B*X = \frac{1}{2 \pi
  i}\left(\Partial{x} + \frac{y}{2} \Partial{s}\right)B; \\  
Y*B = \frac{1}{2 \pi i}\left(\Partial{y} + \frac{x}{2}
  \Partial{s}\right)B, & \hspace{1cm} & B*Y = \frac{1}{2 \pi
  i}\left(\Partial{y} - \frac{x}{2} \Partial{s}\right)B .
\end{eqnarray*}

Consider the action of $\Heisn$ on $\hilbh$ by
\begin{equation} \nonumber
\zeta_{(r,q,p)} v(s,x,y) = e^{-2 \pi i r s} e^{- 2 \pi i \antid (-p X  + q Y)} v(s,x,y),
\end{equation}
where $e^X$ is exponential of the operator of convolution by $X$. 
The elements $(r,0,0)$ act trivially in the representation, $\zeta$,
thus the essential part of the operator $\zeta_{(r,q,p)}$ is
determined by $(q,p)$. If we apply this representation with $r=0$ to
$\voo$ we get a system of vectors $\vab$, 
\begin{equation} \nonumber
v_{(h,q,p)} (s,x,y) = \zeta_{(0,q,p)} \left( \left( \frac{h}{2}
  \right)^n   \exp \left( 2 \pi ihs - \frac{ \pi h}{2} \left(
      \frac{x^2}{\omega m} + y^2 \omega m \right) \right) \right). 
\end{equation}

By (\ref{eq:sametimeevolofkernelandvector}) the vectors $\vab$ are
equivalent to the kernels $\lab$
\begin{equation} \nonumber
\lab = e^{2\pi i (-p \ub{X}{\cdot}{} + q \ub{Y}{\cdot}{} )} \loo.
\end{equation}
Since for any function or distribution, $B$, on $\Heisn$
\begin{equation} \nonumber
\ub{ -p X + q Y  }{B} = -(p x + q y )B
\end{equation}
we have
\begin{equation} \nonumber
\lab =  \exp \left( -2\pi i (qx+py) -2\pi ihs -\frac{\pi h}{2}
  \left( \frac{x^2}{\omega m} +\omega m y^2 \right) \right). 
\end{equation}
\begin{defn}
\textup{\cite{Brodlie03a}}
For $h \in \Space{R}{} \setminus \{ 0 \}$ and $(q,p) \in
\Space{R}{2n}$ define the system of coherent states $k_{(h,q,p)}$ by 
\begin{equation} \nonumber
k_{(h,q,p)} (B) = \langle B* \vab,\vab \rangle = \int_{\Heisn} B(g) \, \lab (g) dg
\end{equation}
\end{defn}
It is clear that the limit as $h \rightarrow 0$ of the kernels $\lab$
will just be the kernels  $\labo$. This proves that the system of
coherent states we have constructed have the $(q,p)$ pure states,
$k_{(0,q,p)}$, from equation (\ref{eq:pclasspurestates}) , as their
limit as $h \rightarrow 0$, which is the content of the next Theorem. 
\begin{thm}
\textup{\cite{Brodlie03a}}
If we have any \(p\)-observable $B$ which is of the form $\delta(s) \hat{F}(x,y)
\newline
$(that is, $B$ is the \(p\)-mechanisation of $F$ see
equations~\eqref{eq:inv-classical-observables} and
~\eqref{eq:p-mechanisation}) then 
\begin{equation} \nonumber
\lim_{h \rightarrow 0} k{(h,q,p)} (B) = k{(0,q,p)} (B) = F(q,p)
\end{equation}
\end{thm}
We have used \(p\)-mechanics to rigorously prove, in a simpler way to
previous attempts \cite{Hepp74}, the classical limit of coherent
states. 
\begin{rem}
If we apply the unitary transformation $\tilde{\alpha}_A$ (from
subsection \ref{sec:quant-from-sypl}) for some $A \in \Symp{n}$ to
some kernel of a $(q,p)$ coherent state, $l_{(0,q,p)}$, we will get
another $(q,p)$ coherent state, $l_{(0,A^t(q,p))}$. 
\begin{equation} \nonumber
l_{(0,q,p)} (s,A (x,y)) = l_{(0,A^t (q,p))} (s,x,y).
\end{equation}
\end{rem}
\section{Examples} \label{sect:examplesjointp}

We now demonstrate the theory through applying it to two examples: the
forced and unforced harmonic oscillator.

\subsection{Unforced Harmonic Oscillator}
\label{ex:harmonic-oscillator} For one account of the unforced
harmonic oscillator see \cite{Kisil00a}, the account we give here is
slightly different.

\begin{figure}[tbp]
  \begin{center}
    \includegraphics{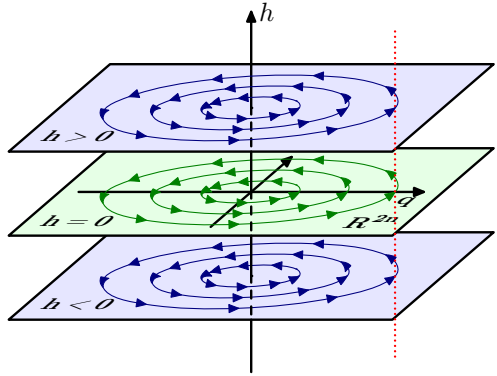}
  \end{center}
    \caption[Dynamics of the harmonic oscillator]{Dynamics of the
      harmonic oscillator in the adjoint space \(\algebra{h}_n^*\) is
      given by the identical linear symplectomorphisms of all orbits
      \(\orbit{\myhbar}\) and \(\orbit{0}\). The vertical dotted
      string is uniformly rotating in the ``horizontal'' plane around
      the \(h\)-axis without any dynamics along the ``vertical''
      direction.}
  \label{fi:harmonic-scillator}
\end{figure}

  Let the \(p\)-mechanical energy function of a harmonic oscillator be
  as obtained in Examples~\ref{ex:harmonic-oscillator-energy}
  and~\ref{ex:quantum-to-p-mechanics}:
  \begin{equation}\label{[eq:H-one]}
    B_H (s,x,y)= -\frac{1}{8\pi^2} \left(m \omega^2
    \delta(s)\delta^{(2)}(x)\delta(y)
      +\frac{1}{m} \delta(s)\delta(x)\delta^{(2)}(y)\right),
  \end{equation} Then the \(p\)-dynamic equation~\eqref{eq:p-equation}
  on \(\Space{H}{n}\) is
  \begin{equation}
    \label{eq:p-oscillator} \frac{d}{dt} B(t;s,x,y)= \sum_{j=1}^n
    \left(\frac{1}{m} x_j \frac{\partial}{\partial y_j} -
      m \omega^2 y_j\frac{\partial}{\partial x_j}\right) B(t;s,x,y).
  \end{equation} Solutions to the above equations are well known to be
  rotations in each of the \((x_j,y_j)\) planes given by:
  \begin{equation}
    \label{eq:ho-evolution} B(t;s,x,y)  =  B_0\left(s, x\cos(\omega
    t) -
      m \omega y\sin( \omega t), 
      \frac{x}{m\omega} \sin(\omega t) + y\cos (\omega t)\right).
  \end{equation}
Since the dynamics on \(\FSpace{L}{2}(\Space{H}{n})\) is given by a
symplectic linear
  transformation of \(\Space{H}{n}\) its Fourier
  transform~\eqref{eq:fourier-transform} to
  \(\FSpace{L}{2}(\algebra{h}_n^*)\) will be the adjoint symplectic
  linear transformations of orbits \(\orbit{\myhbar}\) and
  \(\orbit{0}\) in \(\algebra{h}_n^*\), see
  Figure~\ref{fi:harmonic-scillator}.

  The representations \(\uir{\myhbar}\) transform the energy function
  \(B_H\)~\eqref{[eq:H-one]} into the operator
  \begin{equation}
    \label{eq:quantum-hamiltonian} H_\myhbar
    =-\frac{1}{8\pi^2}(m\omega^2Q^2+\frac{1}{m}P^2),\quad
  \end{equation} where \( Q=d\uir{\myhbar}(X)\) and
  \(P=d\uir{\myhbar}(Y)\) are defined
  in~\eqref{eq:der-repr-h-bar}. The representation \(\uir{(q,p)}\)
  transforms \(B_H\) into the classical Hamiltonian
  \begin{equation}
    \label{eq:classical-hamilt} H (q,p)=\frac{m\omega^2}{2}
    q^2+\frac{1}{2m} p^2.
  \end{equation}

  The \(p\)-dynamic equation \eqref{eq:p-equation} in form
  \eqref{eq:p-oscillator} is transformed by the representations
  \(\uir{\myhbar}\) into the Heisenberg equation
  \begin{equation}
    \label{eq:heisenberg-equation} \frac{d}{dt}
    f(t;Q,P)=\frac{1}{i\hbar}[f,H_\myhbar ], \qquad \textrm{ where }
    \quad \frac{1}{i\hbar}[f,H_\myhbar ]=m\omega^2p\frac{\partial
      f}{\partial q}-\frac{1}{m} q\frac{\partial f}{\partial p},
  \end{equation} defined by the operator
  \(H_\myhbar\)~\eqref{eq:quantum-hamiltonian}. The representation
  \(\uir{(q,p)}\) produces the Hamilton equation
  \begin{equation}
    \label{eq:hamilton-equation} \frac{d}{dt} f(t;q,p)= m \omega^2 p
    \frac{\partial f}{\partial q}- \frac{1}{m} q \frac{\partial
    f}{\partial p}
  \end{equation} defined by the Hamiltonian
  \(H(q,p)\)~\eqref{eq:classical-hamilt}. Finally, to get the solution
  for equations~\eqref{eq:heisenberg-equation}
  and~\eqref{eq:hamilton-equation} it is enough to apply
  representations \(\uir{\myhbar}\) and \(\uir{(q,p)}\) to the
  solution~\eqref{eq:ho-evolution} of the \(p\)-dynamic
  equation~\eqref{eq:p-oscillator}.
To conclude our description of the unforced harmonic oscillator we
give an alternative form of the Hamiltonian which will be of use when
considering the forced harmonic oscillator.
\begin{defn}
\textup{\cite{Brodlie03a}}
We define the \(p\)-mechanical creation and annihilation operators
respectively as convolution by the following distributions 
\begin{eqnarray} \label{eq:defaplus} 
a^+ &=& \frac{1}{2\pi i} (m \omega \zerodelxone - i  \zerodelyone), \\ \label{eq:defaminus}
a^- &=& \frac{1}{2 \pi i} (m \omega \zerodelxone + i  \zerodelyone).
\end{eqnarray}
\end{defn}
The \(p\)-mechanical harmonic oscillator Hamiltonian has the equivalent form
\begin{equation} \nonumber
B_H = \frac{1}{2m} (a^+ * a^- + i \omega m^2 \zerodelsone ).
\end{equation}
We denote the \(p\)-mechanical normalised eigenfunctions of the
harmonic oscillator by $v_n \in \hilbh$ (note here that
$v_0=v_{(h,0,0)}$); they have the form 
\begin{eqnarray} \nonumber
v_n &=& \left( \frac{1}{n!} \right)^{1/2} (\antid a^+)^n * v_{(h,0,0)} \\ \nonumber
&=& \left( \frac{1}{n!} \right)^{1/2} \left( \frac{h}{2} \right)^n  e^{2\pi i hs} (x+i\omega m y)^n \exp \left( \frac{- \pi h}{2} \left( \frac{x^2}{\omega m} + y^2 \omega m \right) \right).
\end{eqnarray}
It can be shown by a trivial calculation that these creation and
annihilation operators raise and lower the eigenfunctions of the
harmonic oscillator respectively. It is important to note that these
states are orthogonal under the $\hilbh$ inner product defined in
equation (\ref{hhip}). 

\subsection[The $p$-Mechanical Forced Oscillator]{The $p$-Mechanical
Forced Oscillator: The Solution and 
Relation to Classical Mechanics} \label{sect:forcedoscclass}

The classical forced oscillator has been studied in great depth for a
long time --- for a description of this see \cite{Jose98} and
\cite{Goldstein80}. The quantum case has also been heavily researched
--- see for example \cite[Sect 14.6]{Merzbacher70},
\cite{Martinez83}. Of interest in the quantum case has been the use of
coherent states, this is described in \cite{Perelomov86}. Here we
extend these approaches to give a unified quantum and classical
solution of the problem based on the \(p\)-mechanical approach. In
\cite{Brodlie03a} there is a more in depth description of this example
and a description of the \(p\)-mechanical scattering matrix.

The classical Hamiltonian for a harmonic oscillator of frequency
$\omega$ and mass $m$ being forced by a real function of a real
variable $z(t)$ (measured in units $\frac{ML}{T^2}$) is
\begin{equation} \nonumber
H(t,q,p) = \frac{1}{2} \left( m \omega^2 q^2 + \frac{1}{m} p^2 \right) - z(t) q.
\end{equation}
Then for any observable $f \in C^{\infty} (\Space{R}{2n})$ the dynamic equation is
\begin{eqnarray} \nonumber
\Fracdiffl{f}{t} &=& \{ f,H \} \\ \label{eq:classdyneqnfo}
&=& \frac{p}{m} \Fracpartial{f}{q} - \omega^2 m q \Fracpartial{f}{p} + z(t) \Fracpartial{f}{p}.
\end{eqnarray}
Through the procedure of \(p\)-mechanisation as described in
subsection~\ref{sec:weyl-quantisation} we get the \(p\)-mechanical forced
oscillator Hamiltonian to be 
\begin{eqnarray} \nonumber
B_H (t;s,x,y) &=& -\frac{1}{8 \pi^2} \left( m \omega^2 \zerodelxtwo +
  \frac{1}{m} \zerodelytwo \right) \\ \nonumber 
&& \qquad- \frac{z(t)}{2 \pi i} \zerodelxone.
\end{eqnarray}
From equation (\ref{eq:p-equation}) the dynamic equation for an
arbitrary observable $B$ is 
\begin{equation} \label{eq:pdynamiceqnforforcedosc}
\Fracdiffl{B}{t} = \frac{x}{m} \Fracpartial{B}{y} - \omega^2 m y \Fracpartial{B}{x} - z(t) y B.
\end{equation}
By substituting the following expression into equation
(\ref{eq:pdynamiceqnforforcedosc}) we see that it is a solution of the
\(p\)-dynamic equation 
\begin{eqnarray}
B (t;s,x,y) &=& \exp \left( 2 \pi i \left( \frac{1}{m \omega} \int_{0}^{t}
    z(\tau) \sin (\omega \tau ) \, d\tau X(t) - \int_{0}^{t} z(\tau )
    \cos (\omega \tau ) \, d \tau Y(t) \right) \right)\nonumber  \\  \label{eq:solnofpmfo}
&& \qquad \qquad \times B (0;s,X(t),Y(t)),
\end{eqnarray}
where
\begin{displaymath}
X(t) = x \cos(\omega t) - m \omega y \sin (\omega t), \quad \textrm{
  and } \quad
Y(t) = \frac{x}{m\omega} \sin (\omega t) + y \cos (\omega t).
\end{displaymath}
Let $F (q,p) = \rho_{(q,p)}(B (s,x,y))$ (i.e. $F$ is the classical
observable corresponding to $B$ under the relationship described in
\cite[Sect. 3.3]{Kisil02e}). 
\begin{eqnarray} \nonumber
F(t;q,p)  &=& \int_{\Space{R}{2n+1}} B (t;s,x,y) e^{2\pi i
    (qx+py)} \, ds \, dx \, dy \\ 
\nonumber
&& = F \left( 0; q \cos(\omega t) -\frac{p}{m\omega} \sin (\omega t) + \frac{1}{m \omega} \int_{0}^{t} z(\tau) \sin (\omega \tau ) \, d\tau , \right.
\\ \label{eq:classicalflowfo}
&& \left. \qquad \qquad q m\omega \sin (\omega t) +p \cos (\omega t) - \int_{0}^{t} z(\tau ) \cos (\omega \tau ) \, d \tau \right).
\end{eqnarray}
This flow satisfies the classical dynamic equation (\ref{eq:classdyneqnfo}) for the forced oscillator --- this is shown in \cite{Jose98}.

\subsection{A Periodic Force and Resonance} \label{sect:periodandres}

In classical mechanics the forced oscillator is of particular interest
if we take the external force to be $z(t)=Z_0 \cos(\Omega t)$
\cite{Jose98}, that is the oscillator is being driven by a harmonic
force of constant frequency $\Omega$ and constant amplitude $Z_0$. By
a simple calculation we have these results for $\Omega \neq \omega$
\begin{equation} 
 \label{eq:sinint}
\int_0^t \cos (\Omega \tau) \sin(\omega \tau) \, d \tau
 = \frac{2}{(\Omega^2 -\omega^2 ) } [ \Omega \cos (\Omega t) \cos (\omega t) + \omega \sin (\Omega t) \sin (\omega t) ]
\end{equation}
\begin{equation}
 \label{eq:cosint}
 \int_0^t \cos (\Omega \tau) \cos(\omega \tau) \, d \tau
 = \frac{2}{(\Omega^2 -\omega^2 ) } [ -\Omega \sin (\Omega t) \cos (\omega t) + \omega \cos (\Omega t) \sin (\omega t) ]
\end{equation}

When these are substituted into (\ref{eq:solnofpmfo}) we see that in
\(p\)-mechanics using a periodic force the \(p\)-mechanical solution
is the flow of the unforced oscillator multiplied by an exponential
term which is also periodic. However this exponential term becomes
infinitely large as $\Omega$ comes close to $\omega$. If we substitute
(\ref{eq:sinint}) and (\ref{eq:cosint}) into
(\ref{eq:classicalflowfo}) we obtain a classical flow which is
periodic but with a singularity as $\Omega$ tends toward
$\omega$. These two effects show a correspondence between classical
and \(p\)-mechanics. The integrals have a different form when $\Omega
= \omega$

\begin{eqnarray} \label{eq:sinint2} \int_0^t \cos (\omega \tau)
\sin(\omega \tau) \, d \tau &=& \frac{1-\cos(2\omega t)}{4 \omega} \\
\label{eq:cosint2} \int_0^t \cos (\omega \tau) \cos(\omega \tau) \, d
\tau &=& \frac{t}{2} + \frac{1}{4 \omega} \sin(2\omega t)
\end{eqnarray}
Now when these new values are substituted into the \(p\)-mechanical
solution (\ref{eq:solnofpmfo}) the exponential term will expand
without bound as $t$ becomes large. When (\ref{eq:sinint2}) and
(\ref{eq:cosint2}) are substituted into (\ref{eq:classicalflowfo}) the
classical flow will also expand without bound --- this is the effect
of resonance. 

\subsection{The Interaction Picture of the Forced Oscillator}
\label{sect:interpictofforcedosc}

We now use the interaction picture to get a better description of the
\(p\)-mechanical forced oscillator. In \cite{Brodlie03a} we use a
different approach to the interaction picture using the $\hilbh$
states, here we use the kernels. The \(p\)-mechanical forced
oscillator Hamiltonian has the equivalent form
\begin{equation} \nonumber
B_H = \frac{1}{2m} \left( a^+ * a^- + i \omega m^2 \zerodelsone \right) - \frac{z(t)}{2} (a^- + a^+ )
\end{equation}
($a^+$ and $a^-$ are the distributions defined in equations (\ref{eq:defaplus}) and (\ref{eq:defaminus})).
We now proceed to solve the forced oscillator in \(p\)-mechanics using
the interaction picture with $B_{H_0} = \frac{1}{2m} (a^+ * a^- + i
\omega m^2 \zerodelsone )$ and $B_{H_1} =  - \frac{z(t)}{2} (a^- + a^+
)$. From (\ref{eq:interforkernel}) the interaction states evolve under
the equation 
\begin{equation} \label{eq:fointerhamil}
\Fracdiffl{\tilde{l}}{t} = \ub{C(s,x,y)}{\tilde{l}}{}
\end{equation}
where
\begin{eqnarray} \nonumber
C(s,x,y) &=& \exp \left( -\frac{t}{2m} \antid (a^+ * a^- + i \omega m^2 \zerodelsone ) \right) \\ \nonumber
&& \times \left( - \frac{z(t)}{2} (a^- + a^+ ) \right) \exp \left( \frac{t}{2m} \antid  (a^+ * a^- + i \omega m^2 \zerodelsone ) \right) ,
\end{eqnarray}
where $\tilde{l} = e^{-t \ub{B_0}{\cdot}{}} l$.
\begin{lem} 
\textup{\cite{Brodlie03a}}
\label{Lem:anihilrel}
We have the relations
\begin{eqnarray} \label{eq:firstanihilrel}
\ub{a^+}{a^-} &=& i \omega m \zerodel \\ \label{eq:secondanihilrel}
\ub{a^+}{a^+ * a^-} &=& i \omega m a^+ \\ \label{eq:thirdanihilrel}
\ub{a^-}{a^+ * a^-} &=& -i \omega m a^- .
\end{eqnarray}
\end{lem}
\begin{proof}
Equation (\ref{eq:firstanihilrel}) follows from simple properties of
commutation for convolutions of Dirac delta functions. Equations
(\ref{eq:secondanihilrel}) and (\ref{eq:thirdanihilrel}) follow from
(\ref{eq:firstanihilrel}) and  the fact that $\ub{}{}$ are a
derivation. 
\end{proof}

\begin{lem} 
\label{Lem:expcommidentity} 
\textup{\cite{Brodlie03a}}
If $B_1,B_2$ are functions or
distributions on $\Heisn$ such that $\ub{B_1}{B_2}{} = \gamma
B_2$ where $\gamma$ is a constant then we have
\begin{equation} \label{eq:expcommidentity}
e^{-\antid \lambda B_1} B_2 e^{\antid \lambda B_1} = e^{-\lambda \gamma} B_2.
\end{equation}
Here $e^{\antid \lambda B_1}$ is the exponential of the operator of
convolution by $\antid \lambda B_1$. 

\end{lem}
The combination of Lemmas \ref{Lem:anihilrel} and
\ref{Lem:expcommidentity} simplifies equation (\ref{eq:fointerhamil})
to 
\begin{eqnarray} \nonumber
\Fracdiffl{\tilde{l}}{t} &=& -\ub{\frac{z(t)}{2} (a^- e^{-i\omega t} +
  a^+ e^{i \omega t})}{ \tilde{l}}{} \\ \nonumber 
&=& \ub{z(t) \cos(\omega t) X - z(t) \sin(\omega t)Y}{\tilde{l}}{}
\end{eqnarray}
where $X$ and $Y$ are from equations \eqref{eq:p-mech-q} and
\eqref{eq:p-mech-p} respectively. This simplifies to 
\begin{equation} \nonumber
\Fracdiffl{\tilde{l}}{t} = 2\pi i (z(t) \cos ( \omega t) x + z(t) sin (\omega t) y ) \tilde{l}
\end{equation}
from which it follows that
\begin{displaymath}
\tilde{l} (t_2,s,x,y) = \exp \left( 2\pi i \left( \int_{t_1}^{t_2}
    z(\tau) \cos (\omega \tau) d\tau x +\int_{t_1}^{t_2} z(\tau) \sin
    (\omega \tau) d\tau y \right) \right) \tilde{l} (t_1,s,x,y). 
\end{displaymath}
 If $\tilde{l} (t_1,s,x,y) = l_{(q,p)}(s,x,y)$ then $\tilde{l}
 (t_2,s,x,y) = \tilde{l}_{(q+\alpha,p+\beta)}(s,x,y)$ where $\alpha =
 \int_{t_1}^{t_2} z(\tau) \cos (\omega \tau) \, d\tau$ and $\beta =
 \int_{t_1}^{t_2} z(\tau) \sin (\omega \tau) \, d\tau$. So if the
 system starts in a coherent state it will remain in a coherent sate
 as time evolves. This result has been found in a much simpler manner
 than the method used in \cite[Sect. 14.6]{Merzbacher70}.

\section{In Conclusion: $p$-Mechanics and Contextuality} 
\label{sec:conclusions}

The presented construction of observables as (convolution) operators
on \(\FSpace{L}{2}(\Space{H}{n})\) and states as positive linear
functionals on them naturally unites the quantum and classical
pictures of mechanics. Moreover the \(p\)-mechanical description of
states through their kernels~\eqref{eq:relationbetweenkernelandvector}
and the Liouville-type equation~\eqref{eq:zerostatestimeeq} for the dynamics
of these kernels is suitable for the \emph{contextual
  interpretation}~\cite{Khrennikov01a,Vaxjo01,Kisil01c} of
quantum mechanics.

Indeed the main point of the contextual
approach~\cite{Khrennikov01a,Vaxjo01} is that in a realistic model the
total probability \(P(E_{12})\) of two disjoint events \(P(E_{1})\) and
\(P(E_{2})\) should not be calculated by a simplistic addition rule
\(P(E_{12})= P(E_{1}) + P(E_{2})\).  The typical
example, when this formula fails, is the two slits
experiment. However the textbook conclusion that ``quantum particles
do not have trajectories'' is not legitimate in the contextual
framework~\cite{KhrenVol01,Kisil01c}.  

Contextuality requires that probabilities of events should depend
from the context of experiments. For example, the probabilities of an electron
to pass the first slit could be either   \(P(E_1|S_1)\) or
\(P(E_1|S_{12})\) depending correspondingly from the context \(S_1\)
(only the first slit is open) or \(S_{12}\) (both slits are open). The
similar  notations \(P(E_2|S_1)\) or \(P(E_2|S_{12})\) are used for
the second slit and in general:
\begin{displaymath}
  P(E_1|S_1)\neq P(E_1|S_{12}), \qquad \textrm{ and } \qquad
  P(E_2|S_2)\neq P(E_2|S_{12}),
\end{displaymath}
Then instead of the wrong probabilities addition rule 
\begin{equation}
\label{eq:wrong-context}
P(E_{12}|S_{12})= P(E_{1}|S_{1}) +P(E_{2}|S_{2})
\end{equation}
the  true contextual addition
of probabilities is:
\begin{equation}\label{eq:quant-add-2}
  P(E_{12}|S_{12})= P(E_{1}|S_{12}) + P(E_{2}|S_{12}).
\end{equation}
Using some relations between contextual probabilities
\(P(E_1|S_1)\),   \(P(E_2|S_2)\),  and \(P(E_1|S_{12})\),
\(P(E_2|S_{12})\), which could be derived from a physical
model, one can improve the wrong formula~\eqref{eq:wrong-context} to
the ``quantum addition'' of probabilities:
\begin{equation}
\label{eq:not-wrong-context}
P(E_{12}|S_{12})= P(E_{1}|S_{1}) +P(E_{2}|S_{2})+\alpha
\sqrt{P(E_{1}|S_{1})P(E_{2}|S_{2})}, 
\end{equation}
where \(\alpha\) is a real number. The presence of the square root
in~\eqref{eq:not-wrong-context}  could be motivated by the
consideration of dimensions. 
If \(\modulus{\alpha}\leq 1\) one
can identify \(\alpha=\cos\phi\) for a quantum phase \(\phi\) and this
would be the standard superposition of states in quantum theory.  In
remaining cases  \(\modulus{\alpha}> 1\) the
formula~\eqref{eq:not-wrong-context} represents the hyperbolic version of
quantum theory~\cite{Khrennikov01a}. 

The contextual calculus of probabilities in quantum mechanics does not
require a superposition of states as linear combinations of vectors
in Hilbert space. Instead the outcome of combined experiments could be
directly calculated from the contextual probabilities in a way similar
to~\eqref{eq:quant-add-2}. \(p\)-Mechanical
equation~\eqref{eq:zerostatestimeeq} for dynamics of states
(i.e. corresponding contextual probabilities) describes the dynamics
in a way similar to classical statistical mechanics. Therefore the
combination of contextual probabilities and \(p\)-mechanical dynamics
form a reliable model for quantum phenomenons. This combination of two
approaches requires further study.

\bibliographystyle{amsplain}
\bibliography{abbrevmr,akisil,analyse,aphysics}

\newcommand{\noopsort}[1]{} \newcommand{\printfirst}[2]{#1}
  \newcommand{\singleletter}[1]{#1} \newcommand{\switchargs}[2]{#2#1}
  \newcommand{\irm}{\textup{I}} \newcommand{\iirm}{\textup{II}}
  \newcommand{\vrm}{\textup{V}}
  \providecommand{\cprime}{'}\providecommand{\arXiv}[1]{\eprint{http://arXiv.o%
rg/abs/#1}{arXiv:#1}}
\providecommand{\bysame}{\leavevmode\hbox to3em{\hrulefill}\thinspace}
\providecommand{\MR}{\relax\ifhmode\unskip\space\fi MR }
\providecommand{\MRhref}[2]{%
  \href{http://www.ams.org/mathscinet-getitem?mr=#1}{#2}
}
\providecommand{\href}[2]{#2}
\begin{thebibliography}{10}

\bibitem{AliAntGazMue}
S.~Twareque Ali, J.-P. Antoine, J.-P. Gazeau, and U.A. Mueller, \emph{Coherent
  states and their generalizations: {A} mathematical overview.}, Rev. Math.
  Phys. \textbf{7} (1995), no.~7, 1013--1104 ({English}), Zbl \# 837.43014.

\bibitem{Arnold91}
V.~I. Arnol{\cprime}d, \emph{Mathematical methods of classical mechanics},
  Graduate Texts in Mathematics, vol.~60, Springer-Verlag, New York, 1991,
  Translated from the 1974 Russian original by K. Vogtmann and A. Weinstein,
  Corrected reprint of the second (1989) edition. \MR{96c:70001}

\bibitem{Arveson76}
William Arveson, \emph{An invitation to {C}*-algebras}, Springer-Verlag,
  {\noopsort{}}1976.

\bibitem{Berezin72}
F.~A. Berezin, \emph{Covariant and contravariant symbols of operators}, Izv.
  Akad. Nauk SSSR Ser. Mat. \textbf{36} (1972), 1134--1167, Reprinted
  in~\cite[pp.~228--261]{Berezin86}. \MR{50 \#2996}

\bibitem{Berezin86}
\bysame, \emph{Metod vtorichnogo kvantovaniya}, second ed., ``Nauka'', Moscow,
  1986, Edited and with a preface by M. K. Polivanov. \MR{89c:81001}

\bibitem{Berezin75}
Felix~A. Berezin, \emph{Quantization in complex symmetric spaces}, Math.
  USSR-Izv. \textbf{9} (1975), 341--379.

\bibitem{Bohm52b}
David Bohm, \emph{A suggested interpretation of the quantum theory in terms of
  ``hidden'' variables. {I}}, Physical Rev. (2) \textbf{85} (1952), 166--179.
  \MR{13,709i}

\bibitem{Bohm52a}
\bysame, \emph{A suggested interpretation of the quantum theory in terms of
  ``hidden'' variables. {II}}, Physical Rev. (2) \textbf{85}
  (\noopsort{1951}1952), 180--193. \MR{13,710a}

\bibitem{Brodlie03a}
Alastair Brodlie, \emph{Classical and quantum coherent states},  (2003), 20,
  \arXiv{quant-ph/0303142}.

\bibitem{CaroSalcedo99}
J.~Caro and L.~L. Salcedo, \emph{Impediments to mixing classical and quantum
  dynamics}, Phys. Rev. \textbf{A60} (1999), 842--852,
  \arXiv{quant-ph/9812046}.

\bibitem{Dixmier77}
Jacques Dixmier, \emph{${C}\sp*$-algebras}, North-Holland Publishing Co.,
  Amsterdam, 1977, Translated from the French by Francis Jellett, North-Holland
  Mathematical Library, Vol. 15. \MR{56 \#16388}

\bibitem{Folland89}
Gerald~B. Folland, \emph{Harmonic analysis in phase space}, Annals of
  Mathematics Studies, vol. 122, Princeton University Press, Princeton, NJ,
  1989. \MR{92k:22017}

\bibitem{Goldstein80}
Herbert Goldstein, \emph{Classical mechanics}, second ed., Addison-Wesley
  Publishing Co., Reading, Mass., 1980, Addison-Wesley Series in Physics.
  \MR{81j:70001}

\bibitem{Hepp74}
K.~Hepp, \emph{The classical limit for quantum mechanical correlation
  functions}, Comm. Math. Phys. \textbf{35} (1974), 265--277.

\bibitem{Honerkamp98}
Josef Honerkamp, \emph{Statistical physics}, Springer-Verlag, Berlin, 1998, An
  advanced approach with applications, Translated from the German manuscript by
  Thomas Filk. \MR{99j:82001}

\bibitem{Howe80b}
Roger Howe, \emph{Quantum mechanics and partial differential equations}, J.
  Funct. Anal. \textbf{38} (1980), no.~2, 188--254. \MR{83b:35166}

\bibitem{Jose98}
Jorge~V. Jos{\'e} and Eugene~J. Saletan, \emph{Classical dynamics}, Cambridge
  University Press, Cambridge, 1998, A contemporary approach. \MR{99g:70001}

\bibitem{Khrennikov01a}
Andrei Khrennikov, \emph{`{Quantum} probabilities' as context depending
  probabilities},  (2001), \arXiv{quant-ph/0106073}.

\bibitem{Vaxjo01}
Andrei Khrennikov (ed.), \emph{Quantum theory: Reconsideration of foundations},
  Mathematical Modelling in Physics, Engineering and Cognitive Science, vol.~2,
  V\"axj\"o University Press, 2002.

\bibitem{KhrenVol01}
A.Yu. Khrennikov and Ya.I. Volovich, \emph{Numerical experiment on interference
  for macroscopic particles},  (2001), \arXiv{quant-ph/0111159}.

\bibitem{Kirillov76}
A.~A. Kirillov, \emph{Elements of the theory of representations},
  Springer-Verlag, Berlin, 1976, Translated from the Russian by Edwin Hewitt,
  Grundlehren der Mathematischen Wissenschaften, Band 220. \MR{54 \#447}

\bibitem{Kirillov94a}
\bysame, \emph{Introduction to the theory of representations and noncommutative
  harmonic analysis [{\MR{90a:22005}}]}, Representation Theory and
  Noncommutative Harmonic Analysis, I, Springer, Berlin, 1994, \MR{1311488}.,
  pp.~1--156, 227--234. \MR{1 311 488}

\bibitem{Kirillov99}
\bysame, \emph{Merits and demerits of the orbit method}, Bull. Amer. Math. Soc.
  (N.S.) \textbf{36} (1999), no.~4, 433--488. \MR{2000h:22001}

\bibitem{Kisil96a}
Vladimir~V. Kisil, \emph{Plain mechanics: Classical and quantum}, J. Natur.
  Geom. \textbf{9} (1996), no.~1, 1--14, \MR{96m:81112}.
  \arXiv{funct-an/9405002}.

\bibitem{Kisil98a}
\bysame, \emph{Wavelets in {Banach} spaces}, Acta Appl. Math. \textbf{59}
  (1999), no.~1, 79--109, \arXiv{math/9807141}. \MR{2001c:43013}.

\bibitem{Kisil02e}
\bysame, \emph{$p$-{Mechanics} as a physical theory. {An} introduction},
  (2002), pp.~25, \arXiv{quant-ph/0212101}.

\bibitem{Kisil00a}
\bysame, \emph{Quantum and classical brackets}, Internat. J. Theoret. Phys.
  \textbf{41} (2002), no.~1, 63--77, \arXiv{math-ph/0007030}. \MR{2003b:81105}

\bibitem{Kisil01c}
\bysame, \emph{Two slits interference is compatible with particles'
  trajectories}, in Khrennikov \cite{Vaxjo01}, \arXiv{quant-ph/0111094},
  pp.~215--226.

\bibitem{Klauder94b}
John~R. Klauder, \emph{Coherent states and coordinate-free quantization},
  Internat. J. Theoret. Phys. \textbf{33} (1994), no.~3, 509--522.
  \MR{95a:81105}

\bibitem{Kurunoglu62}
Behram Kur{\c{s}}uno{\u{g}}lu, \emph{Modern quantum theory}, W. H. Freeman and
  Co., San Francisco, Calif., 1962. \MR{26 \#3375}

\bibitem{Mackey63}
George~W. Mackey, \emph{Mathematical foundations of quantum mechanics},
  W.A.~Benjamin, Inc., New York, Amsterdam, {\noopsort{}}1963.

\bibitem{Martinez83}
Jos{\'e} Martinez, \emph{Diagrammatic solution of the forced oscillator},
  European J. Phys. \textbf{4} (1983), no.~4, 221--227 (1984). \MR{85d:81028}

\bibitem{Merzbacher70}
Eugen Merzbacher, \emph{Quantum mechanics}, John Wiley \& Sons Inc., New York,
  1970. \MR{41 \#4912}

\bibitem{Perelomov86}
A.~Perelomov, \emph{Generalized coherent states and their applications}, Texts
  and Monographs in Physics, Springer-Verlag, Berlin, 1986. \MR{87m:22035}

\bibitem{Prezhdo-Kisil97}
Oleg~V. Prezhdo and Vladimir~V. Kisil, \emph{Mixing quantum and classical
  mechanics}, Phys. Rev. A (3) \textbf{56} (1997), no.~1, 162--175,
  \MR{99j:81010}. \arXiv{quant-ph/9610016}.

\bibitem{Salcedo96}
L.L. Salcedo, \emph{Absence of classical and quantum mixing}, Phys.~Rev.
  \textbf{A54} (1996), 3657--3660, \arXiv{hep-th/9509089}.

\bibitem{Shubin01}
M.~A. Shubin, \emph{Pseudodifferential operators and spectral theory}, second
  ed., Springer-Verlag, Berlin, 2001, Translated from the 1978 Russian original
  by Stig I. Andersson. \MR{2002d:47073}

\bibitem{MTaylor86}
Michael~E. Taylor, \emph{Noncommutative harmonic analysis}, Mathematical
  Surveys and Monographs, vol.~22, American Mathematical Society, Providence,
  RI, 1986. \MR{88a:22021}

\bibitem{vNeumann55}
John von Newmann, \emph{Mathematical foundations of quantum mechanics},
  Princeton University Press, Princeton, {\noopsort{}}1955.

\bibitem{Woodhouse92}
N.~M.~J. Woodhouse, \emph{Geometric quantization}, second ed., The Clarendon
  Press Oxford University Press, New York, 1992, Oxford Science Publications.
  \MR{94a:58082}

\bibitem{Zachos02a}
Cosmas Zachos, \emph{Deformation quantization: quantum mechanics lives and
  works in phase-space}, Internat. J. Modern Phys. A \textbf{17} (2002), no.~3,
  297--316, \arXiv{hep-th/0110114}. \MR{1 888 937}

\end{thebibliography}

\end{document}